\documentclass{aa}
\usepackage[varg]{txfonts} 
\usepackage{natbib} 
\usepackage{graphicx} 
\usepackage{subcaption} 
\usepackage{wasysym} 
\usepackage{verbatim} 
\usepackage{rotating} 
\usepackage{multirow} 
\usepackage[usenames,dvips]{color}
\usepackage{soul} 

\newcommand{\e}[1]{\times 10^{#1}}

\def\apjl{ApJ}%
\def\apjs{ApJS}%
\def\ao{Appl.~Opt.}%
\def\aap{A\&A}%
\title{A comparative study of Type II-P and II-L supernova rise times as exemplified by the case of LSQ13cuw}

\author{E.E.E. Gall \inst{\ref{inst1}\thanks{E-mail: egall01@qub.ac.uk}}
\and J. Polshaw \inst{\ref{inst1}}  
\and R. Kotak \inst{\ref{inst1}}  
\and A. Jerkstrand \inst{\ref{inst1}}  
\and B. Leibundgut \inst{\ref{inst2}}  
\and D. Rabinowitz \inst{\ref{inst3}}  
\and J. Sollerman \inst{\ref{inst5}}  
\and M. Sullivan \inst{\ref{inst6}}  
\and S.J. Smartt \inst{\ref{inst1}}  
\and J.P. Anderson \inst{\ref{inst7}}  
\and S. Benetti \inst{\ref{inst8}}  
\and C. Baltay \inst{\ref{inst9}}  
\and U. Feindt \inst{\ref{inst10},\ref{inst11}}  
\and M. Fraser \inst{\ref{inst4}}  
\and S. Gonz\'{a}lez-Gait\'{a}n \inst{\ref{inst12},\ref{inst13}}  
\and C. Inserra \inst{\ref{inst1}}  
\and K. Maguire \inst{\ref{inst2}}  
\and R. McKinnon \inst{\ref{inst9}}  
\and S. Valenti \inst{\ref{inst14},\ref{inst15}}  
\and D. Young \inst{\ref{inst1}}  
}

\institute{Astrophysics Research Centre, School of Mathematics and Physics, Queen's University Belfast, Belfast BT7 1NN, UK\label{inst1}
\and ESO, Karl-Schwarzschild-Strasse 2, 85748 Garching, Germany\label{inst2}
\and Center for Astronomy and Astrophysics, Yale University, New Haven, CT, USA\label{inst3}
\and Institute of Astronomy, University of Cambridge, Madingley Road, Cambridge, CB3 0HA, UK \label{inst4}
\and Department of Astronomy, The Oskar Klein Centre, Stockholm University, AlbaNova, 10691 Stockholm, Sweden \label{inst5}
\and School of Physics and Astronomy, University of Southampton, Southampton, SO17 1BJ, UK\label{inst6}
\and European Southern Observatory, Alonso de Cordova 3107, Vitacura, Casilla 19001, Santiago, Chile\label{inst7}
\and INAF Osservatorio Astronomico di Padova, Vicolo dell’Osservatorio 5, 35122 Padova, Italy\label{inst8}
\and Department of Physics, Yale University, New Haven, CT 06250-8121, USA\label{inst9}
\and Institut für Physik, Humboldt-Universität zu Berlin, Newtonstr. 15, 12489 Berlin, Germany\label{inst10}
\and Physikalisches Institut, Universität Bonn, Nußallee 12, 53115 Bonn, Germany\label{inst11}
\and Millennium Institute of Astrophysics, Casilla 36-D, Santiago, Chile\label{inst12}
\and Departamento de Astronom\'{i}a, Universidad de Chile, Camino El Observatorio 1515, Las Condes, Santiago, Chile\label{inst13}
\and Las Cumbres Observatory Global Telescope Network, 6740 Cortona Dr., Suite 102, Goleta, CA 93117, USA\label{inst14}
\and Department of Physics, University of California, Santa Barbara, Broida Hall, Mail Code 9530, Santa Barbara, CA 93106-9530,
USA\label{inst15}
}

\date{Received 11th February 2015 /
Accepted 23rd June 2015 }

\abstract{
We report on our findings based on the analysis of observations of the Type II-L
supernova LSQ13cuw within the framework of currently accepted physical predictions
of core-collapse supernova explosions. LSQ13cuw was discovered within a day of
explosion, hitherto unprecedented for Type II-L supernovae.
This motivated a comparative study of Type II-P and II-L supernovae with relatively
well-constrained explosion epochs and rise times to maximum (optical) light.
From our sample of twenty such events, we find evidence of a positive correlation
between the duration of the rise and the peak brightness. On average, SNe II-L tend
to have brighter peak magnitudes and longer rise times than SNe II-P. However,
this difference is clearest only at the extreme ends of the rise time versus peak
brightness relation.
Using two different analytical models, we performed a parameter study to investigate the 
physical parameters that control the rise time behaviour. In general, the models qualitatively 
reproduce aspects of the observed trends. We find that the brightness of the optical peak 
increases for larger progenitor radii and explosion energies, and decreases for larger masses.
The dependence of the rise time on mass and explosion energy is smaller than the dependence on the progenitor radius. We find no evidence that the progenitors of
SNe II-L have significantly smaller radii than those of SNe II-P.
}


\begin{document}

\maketitle

\section{Introduction}

In spite of the surge in supernova (SN) discoveries over recent years, a number of 
persistent issues have remained unsolved for decades. In order to gain insights from 
even larger samples of SN discoveries that are expected with the next generation of 
transient surveys, alternative approaches may be necessary.
Despite significant expenditure of effort, one issue remains: the lack of a complete and consistent mapping between the evolutionary stage during which a massive star 
explodes, and the type of SN that is observed as a result. 

Most -- if not all -- SNe can be classified as belonging to the Type II or Type I
categories depending, respectively, on the presence or absence of hydrogen in their
spectra. A further set of four (``a'',``b'', ``c'', ``n'') spectroscopically motivated 
classifications provides a grouping into distinct subtypes to denote the presence or
absence of either hydrogen or helium, or both. The ``n'' indicates the presence of 
narrow ($\lesssim 1000$\,km\,s$^{-1}$) hydrogen lines in emission, and is equally 
applicable to Type I and II SNe. Although assigning a particular classification to
rare peculiar SNe or transitional objects may not be trivial, the vast majority of SNe
can be typed on the basis of a single spectrum. 
It has long been recognized \citep{Barbon1979} that two further divisions of Type II SNe 
can be made, based solely on light curve morphology post maximum brightness; these are 
the II-plateau (P) and II-linear (L) varieties. As their names imply, the former displays
an extended period (typically $\sim$100\,d) during which the ($VRI$-band) brightness 
remains constant, usually to within $\lesssim 0.5$ magnitudes; Type II-L SNe decline 
linearly at a rate of a few hundredths of a magnitude per day, before settling onto 
the radioactive tail powered phase.

In the canonical picture \citep[e.g.][]{Nomoto1995}, the observed diversity in the spectroscopic 
and photometric behaviour of core-collapse SNe is explained as a one-parameter sequence 
(II-P $\rightarrow$ II-L $\rightarrow$ IIb $\rightarrow$ Ib $\rightarrow$ Ic) governed primarily 
by the hydrogen envelope mass of the progenitor at the time of explosion. Thus, in a
single-star scenario, the expectation would be for higher mass stars to evolve to
a stage with the thinnest of hydrogen envelopes. This picture is intuitively appealing,
and is borne out by observations of the progenitors of several Type II-P SNe, which have been
shown to have red supergiant progenitors, with masses in the range of $8.5 \lesssim M / M_\odot 
\lesssim 16$ \citep[][and references therein]{Smartt2009a}. Barring some exceptions,
direct detections of the progenitors of other subtypes of core-collapse SNe have remained
elusive.

For stars with massive hydrogen envelopes, the plateau phase is attributed to 
the balance between cooling and recombination in the ejecta \citep{Grassberg1971}.
A recombination wave sets in when the ejecta have cooled to $\sim$5-6\,kK, the
recombination temperature for hydrogen. The cooling/recombination wave
moves in a manner such that a roughly constant amount of internal energy is 
advected through it per unit time, resulting in a constant luminosity. 
When the entire envelope has been traversed, the above no longer holds,
marking the end of the pleateau phase.

Suppressing the cooling/recombination wave results in a linearly declining light
curve \citep{Blinnikov1993} and can be achieved if ejecta densities are sufficiently
low \citep{Grassberg1971}. A natural way of fulfilling this requirement is to 
reduce the hydrogen envelope mass to no more than a few solar masses as has been
shown by several authors \citep[][]{Swartz1991, Blinnikov1993}. However, large
progenitor radii are necessary in order to attain the brightness required by observed
light curves.

Thus, within the framework summarized above, one would expect Type II-L SNe to arise from 
stars that have lost more of their hydrogen envelope than the progenitors of Type II-P SNe, 
thereby giving rise to linearly declining light curves. 
Furthermore, one might expect there to be a continuum of decline rates between these two groups 
of Type II SNe.

Over the decades, several studies have considered this very issue. Earlier work by \citet{Patat1994}, who 
considered a heteregeneous sample of Type II SNe, found the $B$-band light curves of Type II-P and II-L
SNe at epochs of $\lesssim$100\,d to overlap in terms of their decline rates. Since then, larger datasets
of well-sampled lightcurves have become available. However, complete consensus on whether the Type II-P and 
II-L SNe show a truly bimodal distribution in decline rates is not evident. The larger (Type II) samples of 
\citet{Anderson2014a} and \citet{Sanders2014} seem to indicate a continuum of decline rates, while \citet{Arcavi2012} 
and \citet{Faran2014b} suggest that distinct subtypes may be apparent if a judicious choice of 
decline rate post-peak brightness is made ($\gtrsim 0.3 - 0.5$ mag. decline at $\sim$50\,d) i.e., 
similar to that proposed by \citet{Li2011a}.

The evolution of the light curve from explosion to peak brightness contains
information about the size, and therefore the type of star that exploded. During the early radiation-dominated
phase, the luminosity is directly proportional to the progenitor radius \citep[][]{Swartz1991,Blinnikov1993}. 
Thus, considering the shape and the total time of the rise to peak brightness for Type II-P and -L SNe, is a 
potentially fruitful endeavour.  
Indeed, for thermonuclear SNe, the exceptionally early discovery of SN~2011fe, led to the first observational
confirmation of a white-dwarf progenitor \citep{Nugent2011}. 

For almost all types of SNe, such observations are particularly challenging given the combination of faintness
at the time of explosion, and the often rapid rise to peak brightness. Still, the type II-P/L SNe are most amenable 
to such studies, as the $^{56}$Ni-powered phase sets in later than in type I SNe. Nevertheless, the rise time behaviour of Type II-L SNe is largely uncharted territory.

Motivated by the discovery of LSQ13cuw soon after explosion, we embarked upon 
a study of Type II-P/L SNe with well-constrained explosion epochs. In what follows, we consider the photometric
and spectroscopic evolution of LSQ13cuw together with a carefully selected sample of SNe assembled from the
literature. In order to identify global trends, we compare the observations with two well-known and 
physically intuitive analytical models. 

A study addressing similar issues was recently presented by \citet{Gonzalez-Gaitan2015}. Although we refrain from detailed comparisons which are beyond the scope of this paper, we note that in spite of broad agreement, there are key differences in the selection criteria of the SN sample, the approach, and the interpretation. We endeavour to highlight the most significant of these in what follows.

The paper is divided into three relatively self-contained parts: 
LSQ13cuw observations are presented in \S\ref{section:Observations_and_data_reduction}; 
the full sample of SNe and comparison with models forms
\S\ref{section:Results_and_discussion}, while a fuller discussion of the 
analytical light curve models and the influence of key parameters appear in the Appendix.

\section{Observations and data reduction}
\label{section:Observations_and_data_reduction}

Supernova LSQ13cuw was discovered on 2013 October 30, during the course of
the La Silla Quest Supernova Survey \citep[LSQ][]{Baltay2013}. The SN was undetected down to a limiting $r'$-band magnitude of 21.21, two days prior to the discovery. It was spectroscopically classified by the PESSTO survey\footnote{Public ESO Spectroscopic Survey of Transient Objects; www.pessto.org} \citep{Smartt2014} first as a Type I SN at $z\sim$0.25 \citep{2013ATel5596} given the combination of a featureless blue continuum, and a long rise time. About a week later, based on second spectrum, it was reclassified as a Type II-P SN at maximum at $z\sim$0.05 \citep{2013ATel5617}. 
Continued monitoring revealed a type II-L light curve (\S\ref{section:Photometry}). 
As part of the PESSTO follow-up campaign, we obtained 5 spectra ranging from 25 to 84 days after explosion;  
additional imaging up to $\sim$100 days after explosion was acquired from a number of different facilities.

\subsection{Data reduction}
\label{section:Data_reduction}

Optical photometry was obtained primarily with the Optical Wide Field Camera, IO:O, mounted on the 2m Liverpool Telescope (LT; $g'r'i'$ filters), with additional epochs from the Las Cumbres Observatory Global Telescope Network 1m telescope (LCOGT; $g'r'i'$ filters) and the ESO Faint Object Spectrograph and Camera, EFOSC2, mounted on the 3.58m New Technology Telescope (NTT; $V\#$641, $g\#$782, $r\#$784, $i\#$705 filters). 
All data were reduced in the standard fashion using the LT/PESSTO\footnote{The PESSTO pipeline, developed by S. Valenti, comprises a set of python scripts which call {\sc pyraf} tasks to reduce EFOSC2 data.}/LCOGT pipelines, including trimming, bias subtraction, and flat-fielding. When necessary, cosmic rays were removed using the {\sc lacosmic} algorithm \citep{vanDokkum2001}.

Point-spread function (PSF) fitting photometry of LSQ13cuw was carried out on all images using the custom built {\sc SNOoPY}\footnote{SuperNOva PhotometrY, a package for SN photometry implemented in IRAF by E. Cappellaro; http://sngroup.oapd.inaf.it/snoopy.html} 
package within {\sc iraf}\footnote{Image Reduction and Analysis Facility, distributed by the National Optical Astronomy Observatories, which are operated by the Association of Universities for Research in Astronomy, Inc, under contract to the National Science Foundation.}. For the $g'r'i'$-bands, a number of local sequence stars (see Table \ref{table:sequence_stars} in the Appendix) from the SDSS DR9 catalogue\footnote{www.sdss3.org} were used to determine colour terms on photometric nights and zeropoints in order to calibrate the photometry of the SN to the SDSS system. For the NTT $V$-band, stars within standard Landolt fields \citep{Landolt1992a} were used for calibration. We estimated the uncertainties of the PSF-fitting via artificial star experiments. An artificial star of the same magnitude as the SN was placed close to the position of the SN. The magnitude was measured, and the process was repeated for several positions around the SN. The standard deviation of the magnitudes of the artificial star were combined in quadrature with the uncertainty of the PSF-fit and the uncertainty of the photometric zeropoint to give the final uncertainty of the magnitude of the SN.
We note that the NTT filters are not standard Sloan filters. $V\#$641 and $r\#$784 are Bessel filters, while $g\#$782 and $i\#$705 are Gunn filters. These usually differ by about 0.1 to 0.2 mag from the corresponding Sloan filters. We have not attempted a correction as the difference is of the same order of magnitude as the error on the measurement, and affects only one epoch in our light curve.

In addition we analysed pre- and post-explosion images for LSQ, which were taken by the ESO 1.0m Schmidt Telescope. The same PSF-fitting technique as described above was performed on the images. The images were taken with a wideband filter \citep{Baltay2013}, and we calibrated the magnitude of the SN to SDSS $r'$ by determining zeropoints using a local sequence of stars (Table \ref{table:sequence_stars}). However, it was not possible to apply a colour transformation since there are no available images of the SN with other filters. Despite this, it is likely that the relatively large uncertainties of the LSQ magnitudes are greater than any differences a colour transformation would make. In the pre-explosion images we determined limiting magnitudes by adding artificial sources to the images at the position of the SN, which were created by building a PSF from nearby stars. 
The magnitude of the PSF was measured and decreased until it was no longer detectable (to 3\,$\sigma$). 
The magnitude of the faintest (3\,$\sigma$) detection was taken to be the limiting magnitude of the image. 
We found the limits to be +21.21 and +21.19\,mag for images taken $-0.11$\,d and $-2.02$\,d before explosion.

LSQ13cuw was independently observed by the Catalina Real-time Transient Survey \citep[CRTS,][]{Drake2009} with the Catalina Sky Survey (CSS) Schmidt telescope under the Transient ID CSS131110:023957-083124. We have included the reported unfiltered magnitudes in Table \ref{table:photometry}, which summarizes the photometric observations for LSQ13cuw. The CRTS data are in good agreement with our LSQ photometry.

A series of five optical spectra were obtained with the NTT+EFOSC2 (see Table \ref{table:journal_spectra}). The spectra were reduced with the PESSTO pipeline using standard techniques. These included trimming, bias subtraction, flat-fielding, wavelength calibration via arc lamps, and flux calibration via spectrophotometric standard stars. The spectra were additionally corrected for telluric absorption by using a model spectrum of the telluric bands. 

Features attributable to the Na\,{\sc i} doublet from interstellar gas either in the Milky Way, or the host galaxy, are not apparent in the spectra of LSQ13cuw (see Figure \ref{figure:LSQ13cuw_spectra}). In order to derive an upper limit to the equivalent width (EW) of a putative Na\,{\sc i}\,D absorption feature, we constructed a weighted stack of all LSQ13cuw spectra, where the weights reflect the signal-to-noise ratio in the 5400-5700\,{\AA} region. We then created a series of artificial Gaussian profiles centred at 5893\,{\AA} and subtracted these from the stacked spectrum. The FWHM of the Gaussian profiles was set to 18\,{\AA}, which corresponds to our lowest resolution spectrum. We increased the EW of the artificial profile in steps of 0.1\,{\AA} in order to determine the limiting EW at which an artificial line profile as described above would be just detectable in the stacked spectrum. Using this method we derive the upper limit for the EW of the Na\,{\sc i}\,D $\lambda\lambda$\,5890,5896 blend to be $<$ 0.9. Using the relation in Equation 9 of \citet{Poznanski2012}, we find $E(B-V)_{\mathrm{host}} <$ 0.16\,mag.
Applying the relation of \citet{Turatto2003c} we find $E(B-V)_{\mathrm{host}} <$ 0.13\,mag.

Recently, \citet{Phillips2013} argued that the EW of the Na\,{\sc i}\,D absorption is unreliable as a measure of the dust extinction from the host galaxies of Type Ia SNe. Nevertheless, a weak or undetectable Na\,{\sc i}\,D absorption seems to be in agreement with little or no extinction.

Other circumstantial factors point towards negligible host galaxy reddening. LSQ13cuw lies in an apparently isolated region at the edge of its presumed host galaxy which is extremely faint (see Section \ref{Section:host_galaxy}). It is unlikely that this region is a pocket of high extinction in an otherwise unremarkable region. Furthermore, recent studies, albeit on SNe Ia \citep[e.g.][]{Holwerda2015}, have shown that the $A_V$ is positively correlated with the radial position of the SN within the host galaxy. With all possible factors taken into account, we assume no host galaxy extinction for LSQ13cuw, and adopt a value of $E(B-V)$ = 0.023\,mag for the Galactic extinction \cite{Schlafly2011}.

We obtained a redshift of 0.0453 $\pm$ 0.0028 for LSQ13cuw from the Balmer lines present in our highest S/N spectrum (+32\,d). 
Assuming $H_0$ = 72\,km\,s$^{-1}$\,Mpc$^{-1}$ this results in a distance of 189\,Mpc $\pm$ 12\,Mpc and an absolute $r'$-band 
peak magnitude of $-$18.04 $\pm$ 0.17.

\subsection{Host galaxy}
\label{Section:host_galaxy}

\begin{figure}
  \centering
    \includegraphics[width=8cm]{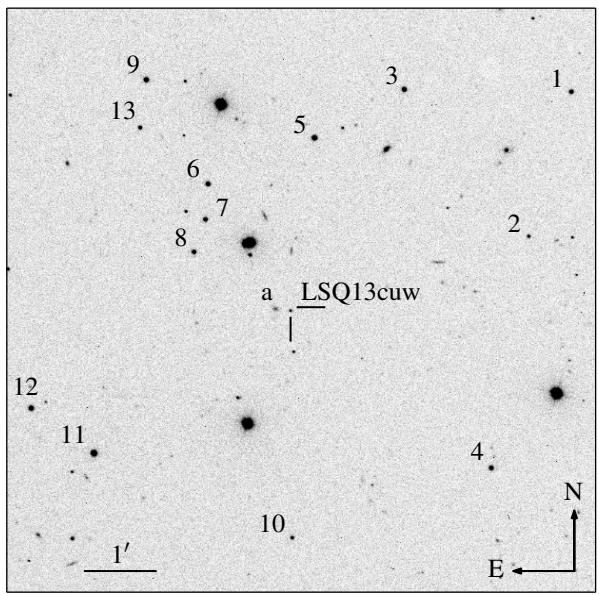}
    \caption{LSQ13cuw and its environment. Short dashes mark the location of the supernova at $\alpha_\mathrm{J2000} = 02^{h}39^{m}57^{s}.35$, 
    $\delta_\mathrm{J2000} = -08^{\circ}31'24''.2$. 
     The nearest galaxy, marked ``a'', originally presumed to be the host, is SDSS J023958.22-083123.5. The numbers mark the positions of the sequence stars (see also Table \ref{table:sequence_stars}) used for the photometric calibrations. $r'$-band image taken on 2013 November 29, 31.6 days after explosion.}
    \label{figure:LSQ13cuw_image}
\end{figure}

\begin{figure}
  \centering
    \includegraphics[width=9cm]{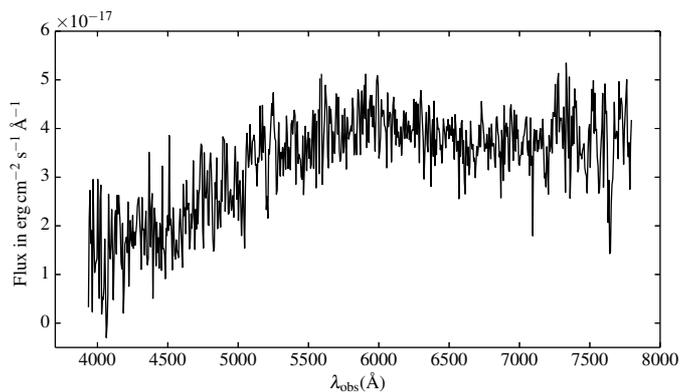}
    \caption{Spectrum of the galaxy SDSS J023958.22-083123.5 designated with ``a'' in Figures \ref{figure:LSQ13cuw_image} and \ref{figure:LSQ13cuw_host}.}
    \label{figure:galaxy}
\end{figure}

LSQ13cuw exploded 13$\farcs$03 W and 0$\farcs$62 S of SDSS J023958.22-083123.5 (Figure \ref{figure:LSQ13cuw_image}). In order to determine whether SDSS J023958.22-083123.5 was indeed the host galaxy of LSQ13cuw, 
we included it in the slit with the SN on the final epoch (2014 Jan. 24) of our spectroscopic observations. The spectrum was reduced in the same way as the SN spectra (see Figure \ref{figure:galaxy} for the reduced and calibrated spectrum). It is almost featureless, so a reliable redshift cannot be determined.
Photometric redshifts from SDSS are $z=0.295 \pm 0.042$ (KD-tree method) and $z=0.219 \pm 0.088$ (RF method\footnote{A description of the KD-tree and the RF method can be found at the following url: https://www.sdss3.org/dr8/algorithms/photo-z.php.}), both being 
significantly higher than the value ($z=0.045$) we obtained directly from spectra of LSQ13cuw. Given the unreliability of photometric redshifts, combined with the fact that the spectrum is reminiscent of an old population, and therefore unlikely to give rise to core-collapse SNe, we deem it improbable that SDSS J023958.22-083123.5 hosted LSQ13cuw. 

\begin{figure}
  \centering
    \includegraphics[width=8cm]{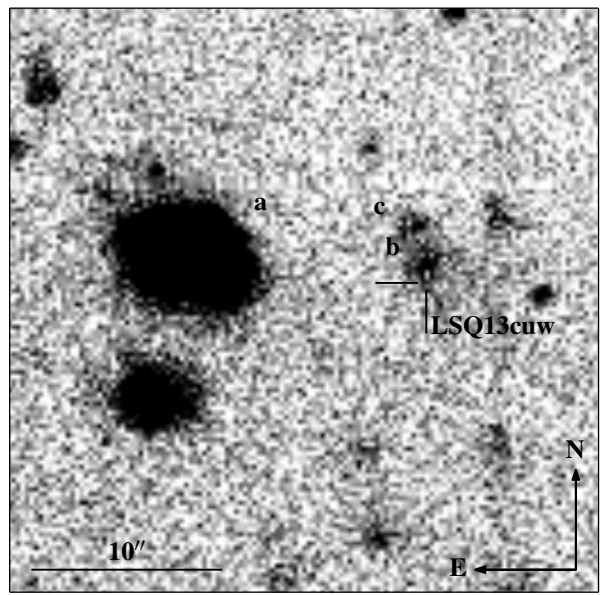}
    \caption{Likely host galaxy of LSQ13cuw. A faint extended source was detected within $0''.5$ of the reported position of SDSS J023957.37-083123.8, $1''.3$ NE from LSQ13cuw. The source appears elongated in the N-S direction with two bright regions at each end that are marked with ``b'' and ``c''. ``a'' marks the position of the galaxy SDSS J023958.22-083123.5 that was originally presumed to be the host. Deep $r'$-band imaging obtained on 2014 September 24, 316.6 days after explosion (rest frame).}
    \label{figure:LSQ13cuw_host}
\end{figure}

In the SDSS-DR10 database, there is a faint extended source, SDSS J023957.37-083123.8, at the SN location. Its photometric redshifts are  0.014 $\pm$ 0.082 (KD-tree method) or 0.049 $\pm$ 0.028 (RF method). The latter would match the 
redshift of the SN within the errors. However, the SDSS DR10 $r'$-band magnitude of SDSS J023957.37-083123.8 ($m_{r'}$ = 20.88 $\pm$ 0.27) is not consistent with 
the limiting magnitude of the LSQ pre-explosion images ($m_{r'} >$ 21.5). Indeed it is flagged as being unreliable, so any agreement
between the SDSS photometric redshift and our own spectroscopically-determined redshift is likely to be purely coincidental.
In order to reliably determine the magnitude of this source, we obtained deep imaging with the NTT ($g$\#782, $r$\#784 and $i$\#705 filters) in September and October, 2014. We took 6 $\times$ 200\,s exposures in each band, and reduced them using the PESSTO pipeline, as detailed in Section \ref{section:Data_reduction}. After aligning and combining the images, we found that the SN had faded beyond detection, and a faint extended source was detected within $0''.5$ of the reported position of SDSS J023957.37-083123.8, $1''.3$ NE from LSQ13cuw. The source appears elongated in the N-S direction with two bright regions at each end (marked with ``b'' and ``c'' in Figure \ref{figure:LSQ13cuw_host}; it is unclear whether the two bright regions are two separate galaxies which are marginally resolved in the images, or are components of a single irregular galaxy). The magnitude was measured using aperture photometry, and an aperture of $3''$ was used which encompasses the entire source (i.e. regions ``b'' and ``c''). The magnitudes of the source are $m_{g’}$ = 22.38 $\pm$ 0.13, $m_{r’}$ = 22.38 $\pm$ 0.13 and $m_{i’}$ = 22.37 $\pm$ 0.32. If the bright region on the north side of the source (``c'' in Figure \ref{figure:LSQ13cuw_host}) is excluded, then the $r'$ and $i'$-band magnitudes are fainter by $\sim$ 0.6 mags. 
The source is unresolved in the $g'$-band. Given the spatial coincidence, we favour SDSS J023957.37-083123.8 as the host galaxy of 
LSQ13cuw, with $M_{r'}$ = $-$14.06 $\pm$ 0.18 (for the entire source ``b'' and ``c''), derived using our redshift estimate above.

\subsection{Photometry}
\label{section:Photometry}

\begin{figure}
  \centering
    \includegraphics[width=9cm]{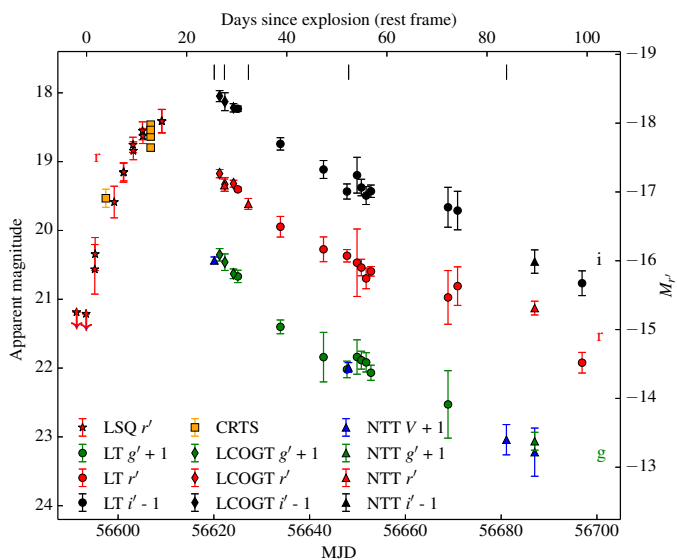}
    \caption{$Vg'r'i'$ light curves of LSQ13cuw. The arrows represent pre-discovery limits. The vertical lines indicate epochs of spectroscopy. We note that the NTT filters are not standard Sloan filters and that CRTS data are unfiltered (Section \ref{section:Data_reduction}).}
    \label{figure:LSQ13cuw_lightcurve}
\end{figure}

Table \ref{table:photometry} shows the log of imaging observations.
\begin{sidewaystable*}
  \caption{Photometry of LSQ13cuw}
  \label{table:photometry}
  \centering
  \begin{tabular}{c c    c            c       c       c               c       c       c               c           c       c               c       c       c               c             }
  \hline               
  Date       & MJD      & Epoch$^*$  &   mag & error & K-corr$^{**}$ & mag   & error & K-corr$^{**}$ & mag       & error & K-corr$^{**}$ & mag   & error & K-corr$^{**}$ & Telescope   \\ 
             &          & rest-frame &   $V$ &       &               & $g'$  &       &               & $r'$      &       &               & $i'$  &       &               & +Instrument \\
\hline                                  
  2013 Oct 26 & 56591.31 &  -2.02     &  -    &  -    &  -            &  -    &  -    &  -            & $>$ 21.19 &  -    &  -            &  -    &  -    &  -            & LSQ ST      \\ 
  2013 Oct 28 & 56593.31 &  -0.11     &  -    &  -    &  -            &  -    &  -    &  -            & $>$ 21.21 &  -    &  -            &  -    &  -    &  -            & LSQ ST      \\ 
  2013 Oct 30 & 56595.13 &   1.64     &  -    &  -    &  -            &  -    &  -    &  -            &     20.57 & 0.36  &  $<$ 0.03     &  -    &  -    &  -            & LSQ ST      \\ 
  2013 Oct 30 & 56595.21 &   1.71     &  -    &  -    &  -            &  -    &  -    &  -            &     20.34 & 0.24  &  $<$ 0.03     &  -    &  -    &  -            & LSQ ST      \\ 
  2013 Nov 01 & 56597.44 &   3.85     &  -    &  -    &  -            &  -    &  -    &  -            &     19.53 & 0.13  &  -            &  -    &  -    &  -            & CRTS CSSST  \\ 
  2013 Nov 03 & 56599.19 &   5.52     &  -    &  -    &  -            &  -    &  -    &  -            &     19.59 & 0.23  &  $<$ 0.03     &  -    &  -    &  -            & LSQ ST      \\ 
  2013 Nov 05 & 56601.11 &   7.36     &  -    &  -    &  -            &  -    &  -    &  -            &     19.16 & 0.14  &  $<$ 0.03     &  -    &  -    &  -            & LSQ ST      \\ 
  2013 Nov 05 & 56601.19 &   7.43     &  -    &  -    &  -            &  -    &  -    &  -            &     19.15 & 0.13  &  $<$ 0.03     &  -    &  -    &  -            & LSQ ST      \\ 
  2013 Nov 07 & 56603.11 &   9.27     &  -    &  -    &  -            &  -    &  -    &  -            &     18.76 & 0.11  &  $<$ 0.03     &  -    &  -    &  -            & LSQ ST      \\ 
  2013 Nov 07 & 56603.19 &   9.35     &  -    &  -    &  -            &  -    &  -    &  -            &     18.84 & 0.13  &  $<$ 0.03     &  -    &  -    &  -            & LSQ ST      \\ 
  2013 Nov 09 & 56605.10 &  11.17     &  -    &  -    &  -            &  -    &  -    &  -            &     18.55 & 0.13  &  $<$ 0.03     &  -    &  -    &  -            & LSQ ST      \\ 
  2013 Nov 09 & 56605.18 &  11.25     &  -    &  -    &  -            &  -    &  -    &  -            &     18.63 & 0.11  &  $<$ 0.03     &  -    &  -    &  -            & LSQ ST      \\ 
  2013 Nov 10 & 56606.79 &  12.79     &  -    &  -    &  -            &  -    &  -    &  -            &     18.64 & 0.03  &  -            &  -    &  -    &  -            & CRTS CSSST  \\ 
  2013 Nov 10 & 56606.80 &  12.80     &  -    &  -    &  -            &  -    &  -    &  -            &     18.54 & 0.03  &  -            &  -    &  -    &  -            & CRTS CSSST  \\ 
  2013 Nov 10 & 56606.80 &  12.80     &  -    &  -    &  -            &  -    &  -    &  -            &     18.80 & 0.03  &  -            &  -    &  -    &  -            & CRTS CSSST  \\ 
  2013 Nov 10 & 56606.81 &  12.81     &  -    &  -    &  -            &  -    &  -    &  -            &     18.46 & 0.03  &  -            &  -    &  -    &  -            & CRTS CSSST  \\ 
  2013 Nov 13 & 56609.09 &  14.99     &  -    &  -    &  -            &  -    &  -    &  -            &     18.41 & 0.17  &  $<$ 0.03     &  -    &  -    &  -            & LSQ ST      \\ 
  2013 Nov 13 & 56609.17 &  15.07     &  -    &  -    &  -            &  -    &  -    &  -            &     18.41 & 0.17  &  $<$ 0.03     &  -    &  -    &  -            & LSQ ST      \\ 
  2013 Nov 24 & 56620.07 &  25.50     & 19.44 & 0.05  & $-$0.02       &  -    &  -    &  -            &      -    &  -    &  -            &  -    &  -    &  -            & NTT+EFOSC2  \\ 
  2013 Nov 25 & 56621.21 &  26.59     &  -    &  -    &  -            & 19.35 & 0.09  &  0.01         &     19.17 & 0.06  &  0.03         & 19.05 & 0.08  &  $-$0.08      & LCOGT       \\ 
  2013 Nov 26 & 56622.19 &  27.52     &  -    &  -    &  -            &  -    &  -    &  -            &     19.34 & 0.09  &  -            &  -    &  -    &  -            & NTT+EFOSC2  \\ 
  2013 Nov 26 & 56622.28 &  27.61     &  -    &  -    &  -            & 19.46 & 0.12  &  0.01         &     19.33 & 0.10  &  0.04         & 19.13 & 0.13  &  $-$0.08      & LCOGT       \\ 
  2013 Nov 28 & 56624.12 &  29.37     &  -    &  -    &  -            & 19.63 & 0.07  &  0.02         &     19.32 & 0.05  &  0.06         & 19.22 & 0.06  &  $-$0.09      & LCOGT       \\ 
  2013 Nov 29 & 56625.03 &  30.24     &  -    &  -    &  -            & 19.67 & 0.09  &  0.02         &     19.40 & 0.03  &  0.06         & 19.23 & 0.04  &  $-$0.10      & LT+IO:O     \\ 
  2013 Dec 01 & 56627.19 &  32.31     &  -    &  -    &  -            &  -    &  -    &  -            &     19.62 & 0.08  &  -            &  -    &  -    &  -            & NTT+EFOSC2  \\ 
  2013 Dec 07 & 56633.90 &  38.73     &  -    &  -    &  -            & 20.40 & 0.10  &  0.03         &     19.95 & 0.15  &  0.13         & 19.74 & 0.09  &  $-$0.15      & LT+IO:O     \\ 
  2013 Dec 16 & 56642.86 &  47.30     &  -    &  -    &  -            & 20.84 & 0.36  &  0.04         &     20.27 & 0.18  &  0.20         & 20.11 & 0.13  &  $-$0.22      & LT+IO:O     \\ 
  2013 Dec 21 & 56647.85 &  52.07     &  -    &  -    &  -            & 21.02 & 0.12  &  0.04         &     20.37 & 0.09  &  0.24         & 20.43 & 0.11  &  $-$0.26      & LT+IO:O     \\ 
  2013 Dec 22 & 56648.12 &  52.33     & 20.99 & 0.07  & $-$0.04       &  -    &  -    &  -            &      -    &  -    &  -            &  -    &  -    &  -            & NTT+EFOSC2  \\ 
  2013 Dec 23 & 56649.94 &  54.07     &  -    &  -    &  -            & 20.84 & 0.25  &  0.04         &     20.47 & 0.49  &  0.26         & 20.20 & 0.26  &  $-$0.28      & LT+IO:O     \\ 
  2013 Dec 24 & 56650.84 &  54.93     &  -    &  -    &  -            & 20.89 & 0.13  &  0.04         &     20.54 & 0.12  &  0.27         & 20.38 & 0.12  &  $-$0.29      & LT+IO:O     \\ 
  2013 Dec 25 & 56651.83 &  55.88     &  -    &  -    &  -            & 20.91 & 0.14  &  0.04         &     20.70 & 0.15  &  0.27         & 20.49 & 0.13  &  $-$0.30      & LT+IO:O     \\ 
  2013 Dec 26 & 56652.84 &  56.84     &  -    &  -    &  -            & 21.07 & 0.11  &  0.05         &     20.59 & 0.07  &  0.28         & 20.43 & 0.09  &  $-$0.31      & LT+IO:O     \\ 
  2014 Jan 11 & 56668.98 &  72.29     &  -    &  -    &  -            & 21.53 & 0.49  &  0.06         &     20.97 & 0.39  &  0.41         & 20.66 & 0.29  &  $-$0.50      & LT+IO:O     \\ 
  2014 Jan 13 & 56670.98 &  74.20     &  -    &  -    &  -            &  -    &  -    &  -            &     20.81 & 0.28  &  0.43         & 20.71 & 0.28  &  $-$0.53      & LT+IO:O     \\ 
  2014 Jan 24 & 56681.08 &  83.86     & 22.04 & 0.22  & $-$0.07       &  -    &  -    &  -            &      -    &  -    &  -            &  -    &  -    &  -            & NTT+EFOSC2  \\ 
  2014 Jan 30 & 56687.08 &  89.60     & 22.22 & 0.35  & $-$0.07       & 22.06 & 0.13  &  -            &     21.13 & 0.10  &  -            & 21.45 & 0.17  &  -            & NTT+EFOSC2  \\ 
  2014 Feb 08 & 56696.85 &  98.95     &  -    &  -    &  -            &  -    &  -    &  -            &     21.92 & 0.15  &  0.63         & 21.77 & 0.18  &  $-$0.95      & LT+IO:O     \\ 
  \hline  
  \end{tabular}
  \\[1.5ex]
  \flushleft
  $^*$The phase is given relative to our estimate of the epoch of explosion on MJD 56593.42. Limiting magnitudes are to 3\,$\sigma$. 
$^{**}$K-corrections were calculated by first using the LSQ13cuw spectroscopy and the {\sc snake} code (SuperNova Algorithm for K-correction Evaluation) within the S3 package (Inserra et al. in prep.). These values were then interpolated to all epochs of photometric observations. The K-corrections are not included in the photometry, as up to an epoch of about 30 days after explosion the derived K-corrections are smaller or similar to the error in the photometry whereas our analysis focusses primarily on the early epochs. 
LSQ = La Silla-QUEST; ST = ESO 1.0m Schmidt Telescope; CRTS = Catalina Real-time Transient Survey; CSSST = Catalina Sky Survey Schmidt telescope; LT = Liverpool Telescope; NTT = New Technology Telescope; LCOGT = Las Cumbres Observatory Global Telescope Network 1m telescope. We note that the NTT filters are not standard Sloan filters and that CRTS data are unfiltered, see also Section \ref{section:Data_reduction}.
The transmission curves for the respective filters are available at: LT $g'r'i'$: http://telescope.livjm.ac.uk/TelInst/Inst/IOO/; LCOGT $g'r'i'$: http://lcogt.net/configdb/filter/; NTT $V\#$641, $g\#$782, $r\#$784, $i\#$705: http://www.eso.org/sci/facilities/lasilla/instruments/efosc/inst/Efosc2Filters.html; LSQ wideband filter: \citet{Baltay2013}, Figure 2.
\end{sidewaystable*}
The light curve is presented in Figure \ref{figure:LSQ13cuw_lightcurve}.

The fact that LSQ13cuw was discovered only two days after the last non-detection allows us to put tight constraints on the explosion epoch. Using these and a low-order polynomial fit to the pre-peak photometry we derive the date of explosion to be MJD 56593.42 $\pm$ 0.68. This makes LSQ13cuw one of the earliest discovered SNe II-L. 

LSQ13cuw rises to peak in 15.9 $\pm$ 1.2\,d (rest frame) with the fitted epoch of maximum light in the $r'$-band being MJD 56610.0 $\pm$ 1.0 and reaching a magnitude of 18.4 $\pm$ 0.1. The $r'$-band light curve then declines linearly until about 26.4\,d (rest frame) after the peak magnitude, then the decline rate slightly slows down. A similar decline to that in the $r'$-band was also observed in the $g'$- and $i'$-bands.

As previously mentioned, the photometric and spectroscopic classifications of SNe are not mutually exclusive.
However, the focus of this work is on explosion-energy driven SNe, so we wish to exclude any objects of 
type Ibc/IIb/IIn even if they have a linearly declining light curve.
Of these, the type IIb SNe are probably the most problematic.
Some SNe IIb show a distinct primary peak, but most do not (e.g. SN 2008ax, \citealt{Pastorello2008,Tsvetkov2009} or SN 2011dh, \citealt{Ergon2014a}). 
As LSQ13cuw was discovered early, it is unlikely that a primary peak was missed. 
We defer a discussion of relative durations of rise times to Section \ref{section:Rise_time_modelling}.

During the post-peak decline phase, the $R$-band light curves of SNe IIb show a similar behaviour to SNe II-L in that they have a relatively linear decline. However, a characteristic property of 
SNe IIb is a very distinct flattening in the blue bands ($U$,$B$,$V$,$g$; see e.g. Figure 2 in \citealt{Ergon2014b}). 

The flattening around 40 days post-explosion in Type IIbs is related to the transition from the $^{56}$Ni-powered diffusion phase to the $^{56}$Ni-powered tail (quasi-nebular) phase. 
In a II-L or II-P SN the light curve in the first $\sim$ 100 days is instead driven by the diffusion of explosion-deposited energy. 

In order to determine whether LSQ13cuw bore more of a resemblance to a type IIb or II-L SN, we measured the decline rates of the Type IIb SNe 1993J \citep{1993IAUC5731,1993IAUC5742,1993IAUC5761,1993IAUC5765,1993IAUC5769,1993IAUC5774,1993IAUC5832,Lewis1994,Barbon1995}, 2008ax, \citep{Pastorello2008,Tsvetkov2009} and 2011dh \citep{Ergon2014a}, of the Type II-L SNe 1979C \citep{Balinskaia1980,deVaucouleurs1981,Barbon1982}, 1990K \citep{Cappellaro1995} and 2001fa \citep{Faran2014b} as well as of LSQ13cuw. 
For the IIb SNe, we measure initial post-peak $V$-band decline rates of 6.4-7.3\,mag/100\,d, which then flatten significantly to declines rates of 1.4-2.2\,mag/100\,d.\footnote{Within the errors, our measurements are consistent with the tail-decline rates of 1.8-1.9\,mag/100\,d in the $V$-band reported by \citet{Ergon2014b} for the three SNe Type IIb 1993J, 2008ax, and 2011dh.}

SN 1979C declines with at a rate of 3.6 $\pm$ 0.2\,mag/100\,d. As it was only discovered relatively late it is not clear whether it also had a somewhat steeper initial decline from maximum. 
SNe 1990K and 2001fa both display a slight shoulder of about 20-30 days in their decline, but then fall even more steeply. SN 1990K has an average decline rate of 4.1 $\pm$ 0.1\,mag/100\,d. SN 2001fa initially declines at a rate of 5.0 $\pm$ 0.4\,mag/100\,d but then flattens to an over all decline rate of 4.1 $\pm$ 0.1\,mag/100\,d.

In contrast, LSQ13cuw initially declines by 5.8 $\pm$ 0.6\,mag/100\,d the $g'$-band and flattens to a decline rate of 3.0 $\pm$ 0.5\,mag/100\,d in the tail. 
Over-all the $g'$-band decline behaviour of LSQ13cuw seems to match the decline rates measured for the Type II-L SNe and we conclude that our classification of LSQ13cuw as a Type II-L SN is justified.

We are aware that both SNe IIb and II-L show a broader range in light curve decline properties than is outlined here, and
other studies previously mentioned above have focussed on the analysis of Type II decline rates.

\subsection{Spectroscopy}
\label{section:Spectroscopy}

\begin{figure}
  \centering
    \includegraphics[width=9cm]{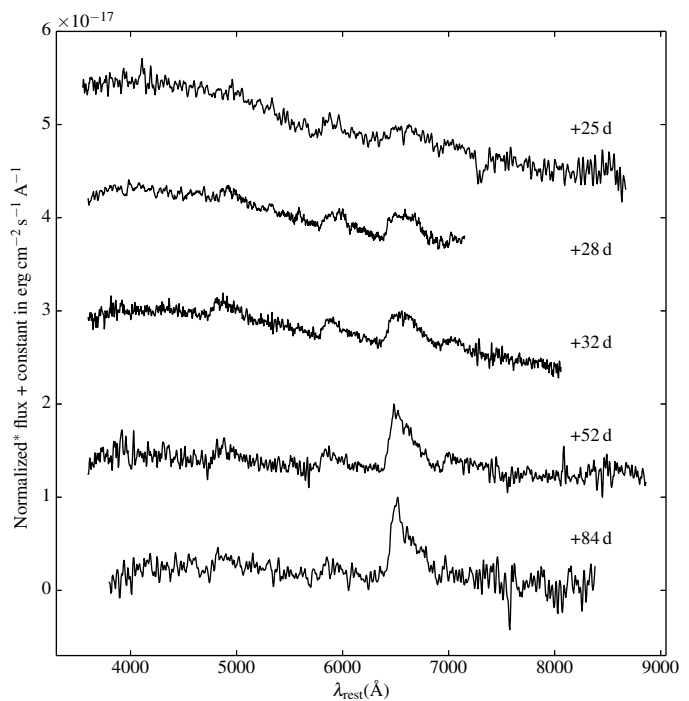}
    \caption{LSQ13cuw spectroscopy. $^*$Flux normalized to the maximum H$\alpha$ flux and smoothed for better visibility of the features. The exact normalizations and smoothing factors are: 
flux/6.9 smoothed by a factor of 3 for the +25\,d spectrum;  
flux/4.3 smoothed by a factor of 5 for +25\,d;
flux/4.3 not smoothed for +32\,d;
flux/4.0 smoothed by a factor of 3 for +52\,d;
flux/2.5 smoothed by a factor of 3 for +84\,d.
}
    \label{figure:LSQ13cuw_spectra}
\end{figure}

Table \ref{table:journal_spectra} shows the journal of spectroscopic observations.
\begin{table*}
  \caption{Journal of spectroscopic observations}
  \label{table:journal_spectra}
  \centering
  \begin{tabular}{c c     c            c              c                      c            }
  \hline               
  Date       & MJD       & Epoch$^*$  & Wavelength   & Telescope            & Resolution \\ 
             &           & rest-frame & range in \AA & +Instrument          &  \AA       \\ 
\hline                
  2013 Nov 24 & 56620.08  & +25.49     & 3550-8670    & NTT+EFOSC2+Gr13      & 18.0       \\
  2013 Nov 26 & 56622.21  & +27.53     & 3600-7150    & NTT+EFOSC2+Gr11      & 13.8       \\
  2013 Dec 01 & 56627.22  & +32.32     & 3600-8060    & NTT+EFOSC2+Gr11,16   & 14.0       \\
  2013 Dec 22 & 56648.15  & +52.33     & 3600-8860    & NTT+EFOSC2+Gr11,16   & 14.0       \\
  2014 Jan 24 & 56681.13  & +83.86     & 3800-8380    & NTT+EFOSC2+Gr13      & 17.7       \\
  \hline  
  \end{tabular}
  \\[1.5ex]
  \flushleft
  $^*$Rest frame epochs (assuming a redshift of 0.0453) with respect to the assumed explosion date of 56593.42 (MJD). The resolution was determined from the FWHM of the sky lines.
\end{table*}
The fully reduced and calibrated spectra of LSQ13cuw are presented in Figure \ref{figure:LSQ13cuw_spectra}. They are corrected for reddening ($E(B - V)$ = 0.023\,mag) and redshift ($z$ = 0.0453). 

The strongest feature in the spectrum is H$\alpha$, which would be matched by the corresponding feature at $\sim$ 4850\,\AA\ in the blue to be H$\beta$. 

In the earliest spectra, both the H$\alpha$ and H$\beta$ profiles are relatively weak but they become stronger at later epochs (see also Table \ref{table:equivalent_width}) and show at best only a very weak absorption component. Recently \citet{Gutierrez2014} analysed a sample of hydrogen-rich Type II SNe and found that the ratio of absorption to emission in the H$\alpha$ P-Cygni profile is smaller for brighter and faster declining light curves and for SNe with higher H$\alpha$ velocities. They attribute this to the mass and density profile of the hydrogen envelope. In this picture we would expect SNe II-L to have smaller H$\alpha$ absorptions than SNe II-P. The fact that LSQ13cuw agrees with these results is encouraging. The sample of SNe II-L presented by \citet{Faran2014b} is also in accord with this conclusion. 

The shape of the H$\alpha$ line profile evolves quite drastically in most Type II SNe. In the earliest spectrum of LSQ13cuw H$\alpha$ is weak, however a few days later the H$\alpha$ line becomes stronger and has a flat top. This might be a possible indication of a detached layer \citep{Jeffery1990}. By +52\,d, the seemingly flat top of the H$\alpha$ feature seen at +32\,d has vanished, leaving a sawtooth shaped profile in its wake. The peak of the H$\alpha$ line is blue-shifted by $\sim$3600\,km\,s$^{-1}$ compared to the rest wavelength. 
Similarly, the H$\alpha$ line in the +84\,d spectrum is blue shifted by $\sim$2000\,km\,s$^{-1}$. 

A blue-shifted peak in the H$\alpha$ profile was already noted for SN 1979C. \citet{Chugai1985} suggest this might be due to an opaque core screening the atmospheric layers in the expanding envelope in which the H$\alpha$ line is formed.
\citet{Anderson2014b} have investigated the occurrence of blue shifted emission peaks in 95 Type II supernovae and come to the conclusion that it is a generic property, observed for the majority of objects in their sample. They perform non-LTE time-dependent radiative-transfer simulations and show that the shape of the H$\alpha$ line profile is a result of the steep density profile of the H layers of the ejecta. Since the opacity is dominated by electron scattering and also the line emissions mainly come from the region below the continuum photosphere the probability of observing blue-shifted line photons is higher. This effect decreases with time as the ejecta expand and the density decreases.

We can also discern a feature at about 5900\,\AA. This is typically attributed to a blend of He\,{\sc i} $\lambda$5876 and Na\,{\sc i}\,D $\lambda\lambda$5890,5896. As with the hydrogen line profiles, it only shows a very weak absorption. At about 7100\,\AA\ there seems to be another emission feature, which shows weakly in the +32\,d spectrum and a bit stronger in the +52\,d spectrum. We suspect this to result from He\,{\sc i} $\lambda$7065.

\begin{table*}
  \caption{Emission equivalent width measurements for LSQ13cuw.}
  \label{table:equivalent_width}
  \centering
  \begin{tabular}{c c     c            c                   c                  c                      }
  \hline                      
  Date       & MJD       & Epoch$^*$  & H$\alpha$         & H$\beta$         & He\,{\sc i}/Na\,{\sc i}\,D$^{**}$ \\ 
             &           & rest-frame & EW in \AA         & EW in \AA        & EW in \AA             \\ 
\hline                          
  2013 Nov 24 & 56620.08  & +25.49     & $-41  \pm 1  $    & $-16  \pm 7 $    & $-34  \pm 4 $         \\
  2013 Nov 26 & 56622.21  & +27.53     & $-109 \pm 10 $    & $-18  \pm 2 $    & $-41  \pm 5 $         \\
  2013 Dec 01 & 56627.22  & +32.32     & $-152 \pm 24 $    & $-57  \pm 8 $    & $-47  \pm 8 $         \\
  2013 Dec 22 & 56648.15  & +52.33     & $-480 \pm 73 $    & $-170 \pm 37$    & $-146 \pm 27$         \\
  2014 Jan 24 & 56681.13  & +83.86     & $-800 \pm 190$    & $-242 \pm 85$    & $-276 \pm 50$         \\
  \hline  
  \end{tabular}
  \\[1.5ex]
  \flushleft
  $^*$Rest frame epochs (assuming a redshift of 0.0453) with respect to the assumed explosion date of 56593.42 (MJD). $^{**}$He\,{\sc i} $\lambda$5876 and Na\,{\sc i}\,D $\lambda\lambda$5890,5896 blend.
\end{table*}

\begin{figure}
  \centering
    \includegraphics[width=9cm]{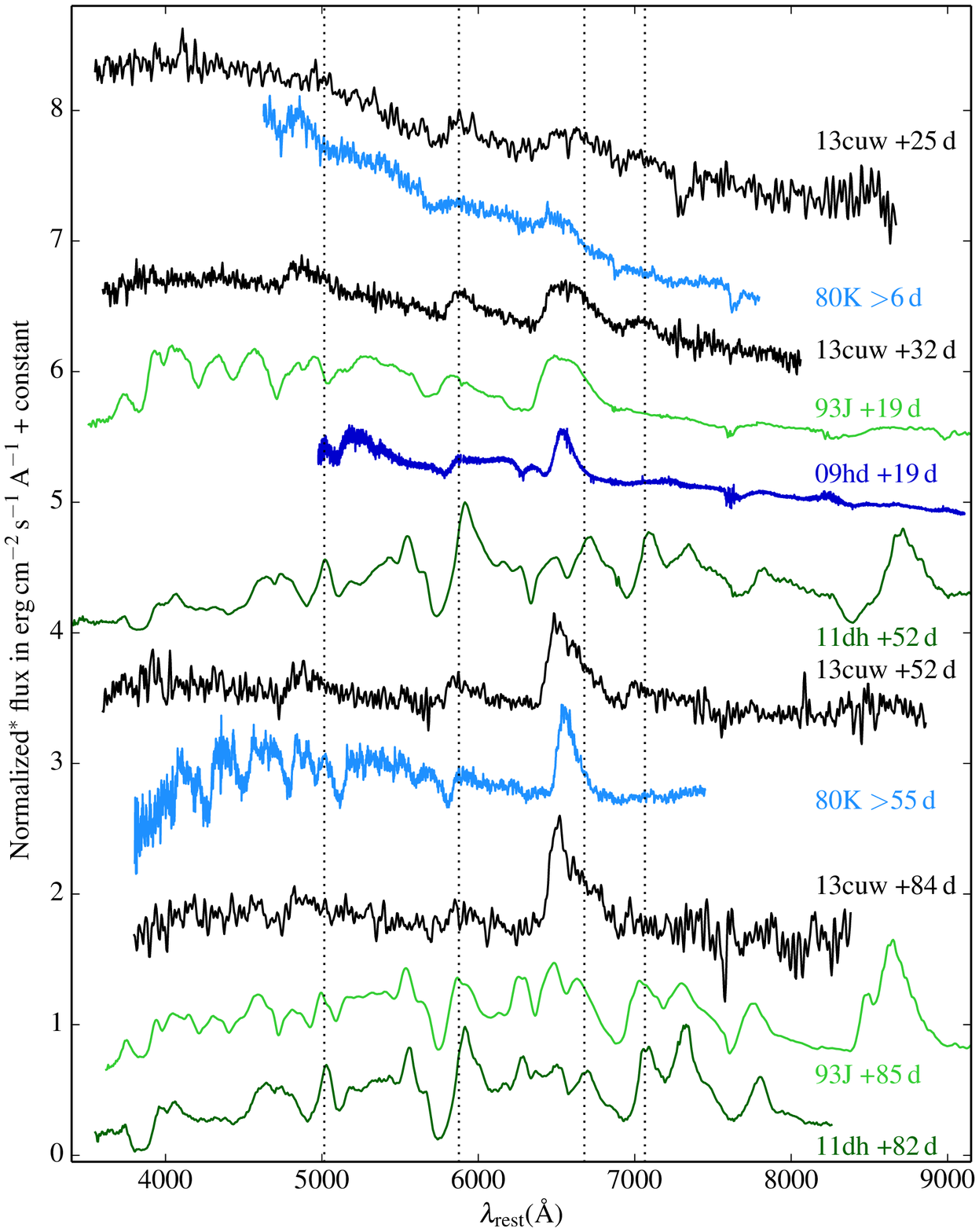}
    \caption{LSQ13cuw spectra in comparison with spectra of the SNe 1980K \citep[II-L;][]{Uomoto1986}, 1993J \citep[IIb;][]{Jeffery1994,Barbon1995,Fransson2005}, 2009hd \citep[II-L;][]{Elias-Rosa2011} and 2011dh \citep[IIb;][]{Ergon2014a}. The epochs are given relative to the estimated date of explosion, with the exception of SN 1980K for which the epochs are given relative to the first detection. The dotted lines correspond to the positions of He\,{\sc i} $\lambda$5015, $\lambda$5876, $\lambda$6678 and $\lambda$7065.
$^*$Flux normalized to the maximum H$\alpha$ flux and smoothed for better visibility of the features. 
}
    \label{figure:LSQ13cuw_spectra_comparison}
\end{figure}

In Figure \ref{figure:LSQ13cuw_spectra_comparison} we compare the spectra of LSQ13cuw with spectra from other Type II SNe. For all SNe the epoch is given relative to the estimated date of explosion. The exception is SN 1980K \citep[e.g.][]{Uomoto1986} for which no constraints on the explosion epoch are available and the epochs are therefore given relative to the first detection. 

In the earliest spectrum of LSQ13cuw we see a few weak features similar to the early spectrum of the Type II-L SN 1980K.  
The flat top of the H$\alpha$ line of LSQ13cuw a few days later has previously also been observed in the early spectra of the Type IIb SN 1993J. However, in the further evolution of SN 1993J the H$\alpha$ emission splits into two components. \citet{Matheson2000a} attributed this to clumping in the ejecta of SN 1993J. We see no evidence of this in the spectra of LSQ13cuw.\footnote{The small absorption in the centre of the H$\alpha$ profile of LSQ13cuw is a telluric feature.} 

The shape of the H$\alpha$ profile is quite similar in the +25 and +32\,d spectra of LSQ13cuw, the +19\,d spectrum of SN 1993J, and the +19\,d spectrum of 2009hd, even though they represent epochs that are two weeks apart. This might indicate that different SNe go through the different evolutionary stages on different time scales. 

The sawtooth shape in the +52\,d spectrum the H$\alpha$ feature in LSQ13cuw strongly resembles the H$\alpha$ profile of the $>$55\,d spectrum of SN 1980K, which is similarly asymmetric. 

Compared to the other Type II-L SNe like 1980K and 2009hd, it seems that in LSQ13cuw the H$\alpha$ feature is generally broader. Only the SN 1993J spectra display similarly broad H$\alpha$ profiles. This might be an indication of a somewhat broader velocity distribution.

Apart from H$\alpha$ we also see H$\beta$ in the various SNe as well as the blend of He\,{\sc i} $\lambda$5876 and Na\,{\sc i}\,D $\lambda\lambda$5890,5896. The He\,{\sc i} $\lambda$7065 line does not show in any of the comparison spectra. The feature between 7000 and 7300\,\AA\ in the late spectrum of SN 1993J is probably due to the [Ca\,{\sc ii}] $\lambda\lambda$7291,7324 doublet rather than He\,{\sc i}. 

Given the presence of a feature near $\lambda$5880\,{\AA}, close to He\,{\sc i} $\lambda$5876 in the 25 and 32\,d spectra
of LSQ13cuw (Figures 5, 6), we must consider whether a spectroscopic classification of Type IIb is warranted.
A closer inspection reveals marked differences to Type IIb spectra (Figure 6). We note that prototypical IIb SNe
such as SNe 1993J and 2011dh show optical signatures of helium about two weeks post-explosion \citep[e.g.][]{Sahu2013} with hydrogen lines becoming progressively weaker with time. In contrast, LSQ13cuw maintains a strong feature due
to H$\alpha$ at epochs as late as 84\,d, while the putative $\lambda$5876 feature is no longer discernible.
Although we cannot rule out the emergence of He\,{\sc i} $\lambda$6678 in the red wing of H$\alpha$ giving rise to
the pronounced asymmetry, the relative strengths of the hydrogen versus helium lines argues against a firm
Type IIb classification. In Section 3.4 we show that the rise time behaviour of {\it bona fide\/} IIb SNe
is also markedly different from Type IIP-L SNe.

Finally, it is interesting to note that compared to the other SNe the spectra of LSQ13cuw show almost no, or only extremely weak lines of intermediate mass elements between 4000 and 5500\,\AA, which we would expect to appear a few weeks after explosion.

\section{Results and discussion}
\label{section:Results_and_discussion}

\subsection{The sample}

The scarcity of SNe II-L poses a challenge in the quest to compare LSQ13cuw to other SNe II-L. There are a few well-known objects in this class, like SN 1979C \citep[e.g.][]{Balinskaia1980,Panagia1980,deVaucouleurs1981,Branch1981,Barbon1982}, SN 1980K \citep[e.g.][]{Buta1982a,Buta1982b,Thompson1982}, SN 1990K \citep{Cappellaro1995} and SN 1996L \citep{Benetti1999}, however while we can use their data to compare the post-maximum behaviour, none of the aforementioned objects was observed during its rise to peak. 

\citet{Faran2014b} recently published a set of eleven Type II-L SNe, three of which have good pre-maximum photometry and a well constrained explosion epoch, that we can therefore add to our sample: SNe 2001fa, 2003hf and 2005dq.

For SNe II-P the situation is considerably better with quite a few SNe II-P having been discovered and observed at early stages: SN 1999em \citep[e.g.][]{Hamuy2001,Leonard2002a,Leonard2002b,Elmhamdi2003}, SN 2004et \citep[e.g.][]{Li2005,Sahu2006,Kotak2009} and SN 2005cs \citep[e.g.][]{Pastorello2006,Tsvetkov2006,Brown2007}, to name a few, but see also Table \ref{table:SN_all}.

Our analysis is mainly performed in the $r'$/$R$-band, where all objects in our sample have reasonably good quality data. In the following, we will refer to the $r'$/$R$-band for all observables unless explicitly stated otherwise. 

\begin{figure*}[t!]
   \centering
   \begin{subfigure}[t]{0.5\textwidth}
      \includegraphics[height=7cm]{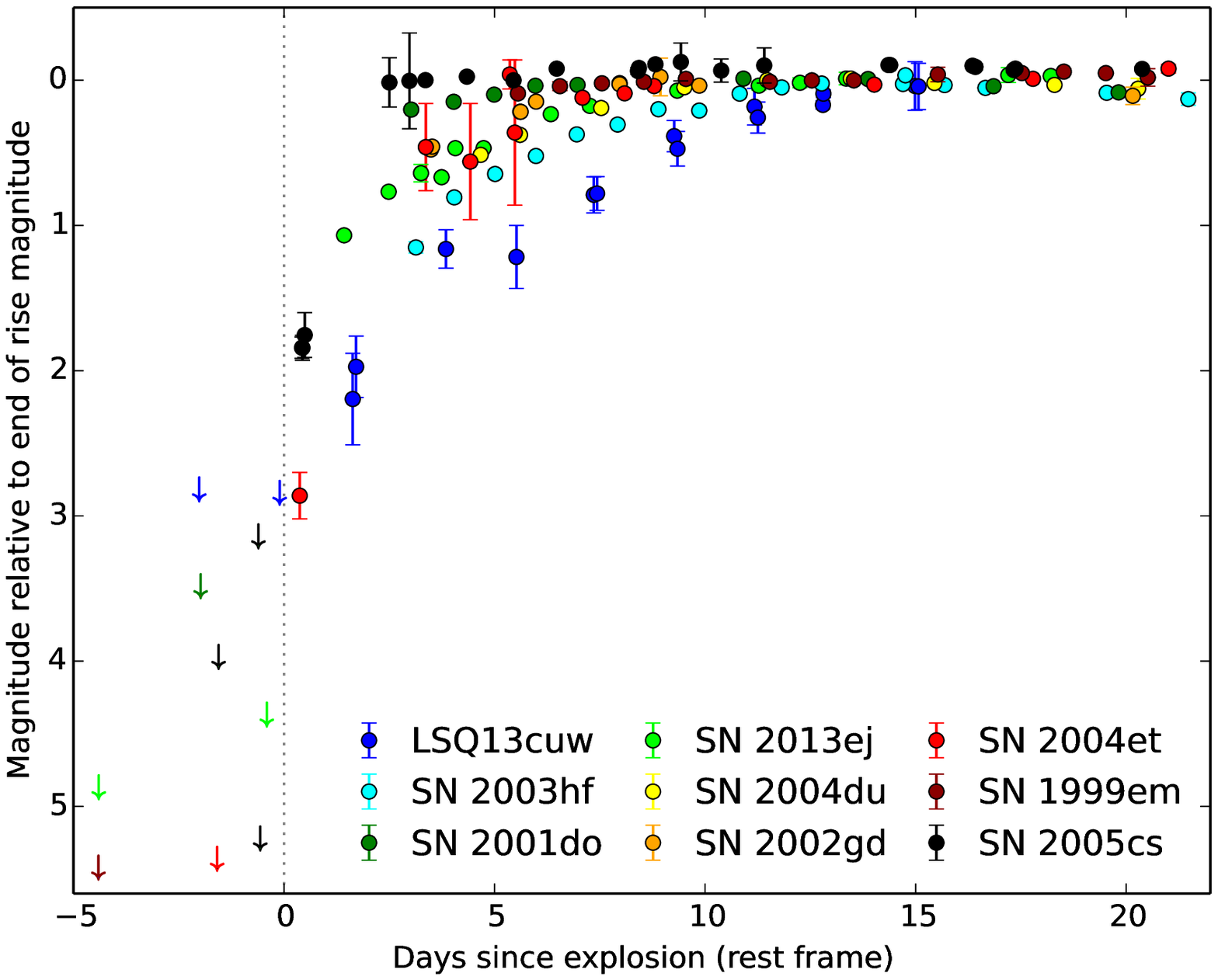}
      \caption{$r'$/$R$-band pre-peak light curves.}
      \label{figure:LC_rise_comparison}
   \end{subfigure}%
   \begin{subfigure}[t]{0.5\textwidth}
      \includegraphics[height=7cm]{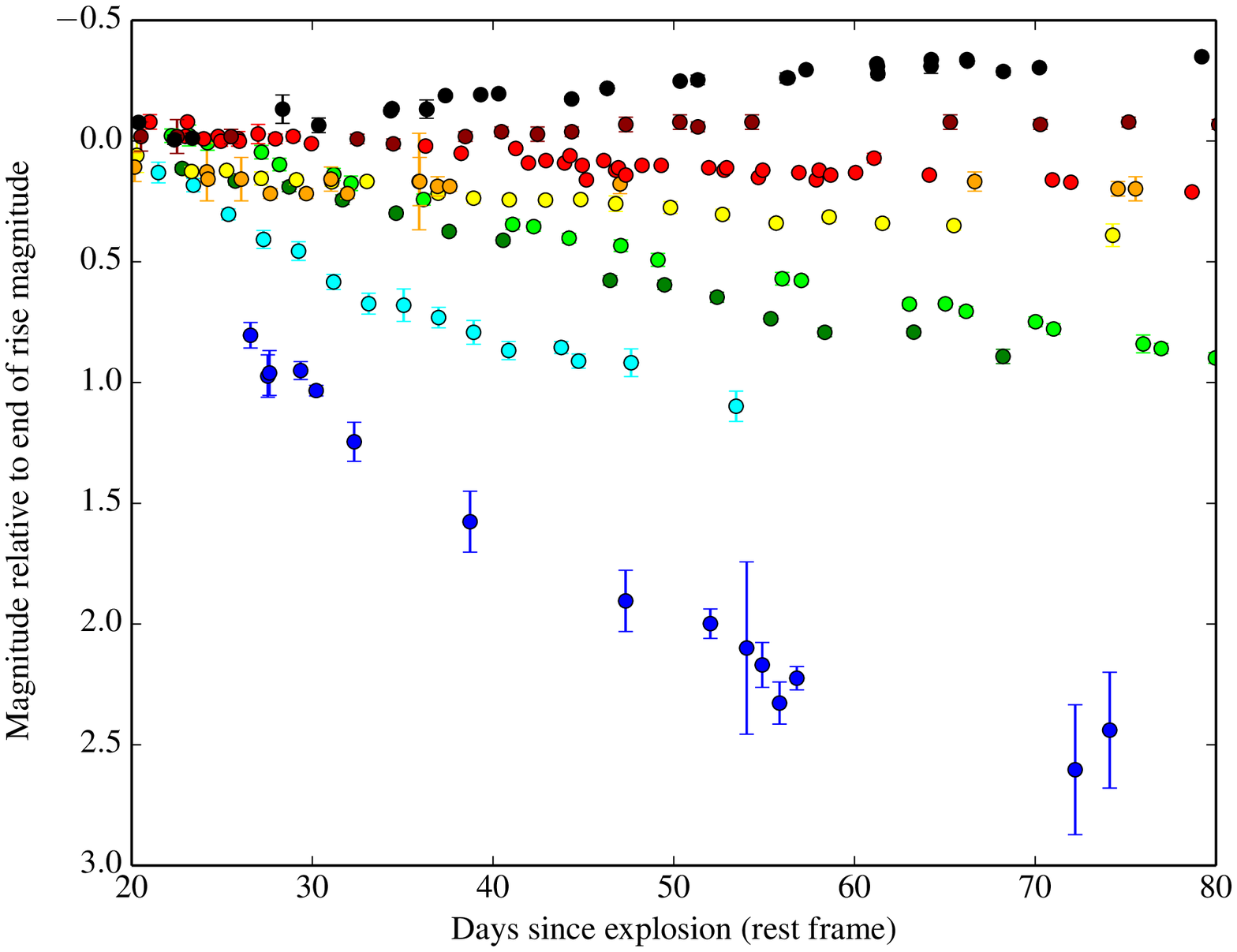}
      \caption{$r'$/$R$-band post-peak light curves.}
      \label{figure:LC_decline_comparison}
   \end{subfigure}%
   \caption{$r'$/$R$-band light curve comparison of a selection of SNe II-P and II-L from our sample. Magnitudes are plotted relative to the ``end-of-rise'' magnitude (see Section \ref{section:Rise_times}) to visualize the range in rise times and decline rates amongst the various SNe. The arrows show the limits on the pre-discovery non-detections. The dotted line marks the explosion epoch.}
   \label{figure:LC_comparison}
\end{figure*}

Our sample and the light curve properties are presented in Table \ref{table:SN_all}. Figure \ref{figure:LC_comparison} shows a comparison of pre- and post-peak light curves for selected SNe from our sample.

\begin{sidewaystable*}
  \caption{Comparison of Type II SN properties}
  \label{table:SN_all}
  \centering
  \begin{tabular}{l c    c                          l                              l                          l                                l@{}        l                             c                     r@{}    l                               c                 }                             
  \hline                              
  SN        & Type$^*$  &  Host galaxy             & \multicolumn{2}{c}{$E(B - V)$}                          & \multicolumn{1}{c}{$\mu^{**}$} & \multicolumn{2}{c}{$t_0$ (MJD)$^{***}$} & End of rise         & \multicolumn{2}{c}{Rise time in}      &                 \\ 
            &           &                          & \multicolumn{1}{c}{Galactic} & \multicolumn{1}{c}{Host} &                                &           &                             & absolute magnitude  & \multicolumn{2}{c}{days (rest frame)} &                 \\
 \hline                              
  SN 1990E  & II-P      &  NGC 1035                & 0.022                        & 0.46 $\pm$ 0.10          & 31.48 $\pm$ 0.22               &  47934.56 & $\ \pm$ 2.56                & $-$17.69 $\pm$ 0.34 &   9.6 & $\ \pm$ 2.6                   & a,b,c,d,e       \\  
  SN 1999em & II-P      &  NGC 1637                & 0.04                         & 0.06                     & 30.23 $\pm$ 0.14               &  51475.87 & $\ \pm$ 2.59                & $-$16.90 $\pm$ 0.14 &   8.3 & $\ \pm$ 2.7                   & a,f,g,h,i       \\ 
  SN 1999gi & II-P      &  NGC 3184                & 0.017                        & 0.19 $\pm$ 0.09          & 30.54 $\pm$ 0.29               &  51514.84 & $\ \pm$ 0.34                & $-$16.75 $\pm$ 0.38 &  11.5 & $\ \pm$ 0.6                   & a,b,j,k,l       \\ 
  SN 2002gd & II-P      &  NGC 7537                & 0.059                        &  -                       & 32.87 $\pm$ 0.35               &  52551.5  & $\ \pm$ 2.0                 & $-$16.10 $\pm$ 0.35 &   8.1 & $\ \pm$ 2.2                   & a,m             \\
  SN 2004du & II-P      &  UGC 11683               & 0.084                        &  -                       & 33.94 $\pm$ 0.13               &  53225.6  & $\ \pm$ 1.1                 & $-$17.67 $\pm$ 0.13 &  10.9 & $\ \pm$ 1.5                   & a,b,n           \\
  SN 2004et & II-P      &  NGC 6946                & 0.34                         & 0.07 $\pm$ 0.07          & 28.37 $\pm$ 0.15               &  53270.61 & $\ \pm$ 0.60                & $-$17.15 $\pm$ 0.24 &   8.6 & $\ \pm$ 0.9                   & a,b,o,p,q,r     \\
  SN 2005cs & II-P      &  NGC 5194                & 0.032                        & 0.08 $\pm$ 0.04          & 29.26 $\pm$ 0.33               &  53548.5  & $\ \pm$ 0.5                 & $-$15.12 $\pm$ 0.35 &   3.1 & $\ \pm$ 0.5                   & a,s,t,u,v,w,x   \\
  SN 2006bp & II-P      &  NGC 3953                & 0.027                        &  -                       & 31.33 $\pm$ 0.19               &  53833.7  & $\ \pm$ 0.3                 & $-$16.46 $\pm$ 0.19 &   6.7 & $\ \pm$ 0.4                   & a,b,w,y         \\ 
  SN 2010id & II-P      &  NGC 7483                & 0.054                        &  -                       & 33.12 $\pm$ 0.48               &  55452.3  & $\ \pm$ 2.0                 & $-$14.37 $\pm$ 0.48 &   6.2 & $\ \pm$ 2.0                   & a,b,z           \\ 
  SN 2012A  & II-P      &  NGC 3239                & 0.029                        &  -                       & 29.96 $\pm$ 0.15               &  55933.0  & $\ ^{+ 1.0}_{- 3.0}$        & $-$16.33 $\pm$ 0.15 &   8.1 & $\ \pm$ 3.0                   & a,A             \\ 
  SN 2000dc & II-P/II-L &  ESO 527-G019            & 0.071                        &  -                       & 32.93 $\pm$ 0.14               &  51761.8  & $\ \pm$ 3.5                 & $-$17.25 $\pm$ 0.19 &  10.5 & $\ \pm$ 4.9                   & a,b,n,B         \\ 
  SN 2001cy & II-P/II-L &  UGC 11927               & 0.185                        &  -                       & 33.01 $\pm$ 0.12               &  52085.0  & $\ \pm$ 5.5                 & $-$17.65 $\pm$ 0.12 &   5.4 & $\ \pm$ 5.5                   & a,b,n,B,C       \\ 
  SN 2001do & II-P/II-L &  UGC 11459               & 0.170                        &  -                       & 32.35 $\pm$ 0.15               &  52133.2  & $\ \pm$ 2.0                 & $-$17.27 $\pm$ 0.15 &   8.1 & $\ \pm$ 2.0                   & a,b,n,B         \\ 
  SN 2013ej & II-P/II-L &  NGC 0628                & 0.061                        &  -                       & 29.77 $\pm$ 0.28               &  56496.95 & $\ \pm$ 0.92                & $-$17.60 $\pm$ 0.28 &  12.9 & $\ \pm$ 1.4                   & a,b,D,E,F,G,H,I \\ 
  SN 1979C  & II-L      &  NGC 4321                & 0.023                        & 0.16 $\pm$ 0.04          & 30.30 $\pm$ 0.15               &  43970    & $\ \pm$ 8                   & $-$18.72 $\pm$ 0.37 &  23.9 & $\ \pm$ 8.0                   & a,b,J,K,L       \\
  SN 2001fa & II-L      &  NGC 0673                & 0.069                        &  -                       & 34.09 $\pm$ 0.27               &  52197.9  & $\ \pm$ 2.5                 & $-$18.13 $\pm$ 0.27 &   7.8 & $\ \pm$ 2.6                   & a,b,B,M         \\
  SN 2003hf & II-L      &  UGC 10586               & 0.019                        &  -                       & 35.55 $\pm$ 0.04               &  52863    & $\ \pm$ 2                   & $-$18.46 $\pm$ 0.05 &  13.2 & $\ \pm$ 2.1                   & a,b,B           \\
  SN 2005dq & II-L      &  UGC 12177               & 0.070                        &  -                       & 34.77 $\pm$ 0.03               &  53608    & $\ \pm$ 4                   & $-$17.50 $\pm$ 0.03 &   9.1 & $\ \pm$ 3.9                   & a,b,B           \\
  SN 2009hd & II-L      &  NGC 3627                & 0.030                        & 1.20                     & 29.86 $\pm$ 0.08               &  55001.6  & $\ \pm$ 3.9                 & $-$17.11 $\pm$ 0.16 &  15.9 & $\ \pm$ 4.1                   & a,b,N,O,P       \\
  LSQ13cuw  & II-L      & SDSS J023957.37-083123.8 & 0.023                        &  -                       & 36.38 $\pm$ 0.13               &  56593.42 & $\ \pm$ 0.68                & $-$18.03 $\pm$ 0.13 &  15.0 & $\ \pm$ 0.8                   & a,b             \\
  \hline  &      
  \end{tabular}
  \\[1.5ex]
  \flushleft
$^*$Type for each SN adopted from literature;
$^{**}H_0$ = 72\,km\,s$^{-1}$\,Mpc$^{-1}$
$^{***}$For most SNe the epoch of explosion was adopted from the literature. In the cases where no estimate was given we derived the explosion epoch by taking the average MJD between the first detection and the last pre-discovery non-detection (SNe 1979C, 2000dc, 2001cy, 1999gn, 1999gi, 2009hd, 2001do, 2001fa). For SNe SN 2004du and 2004et no pre-discovery non-detections were reported, we therefore used a low-order polynomial fit to the observed data to estimate the explosion epoch. For SN 1999em we adopted the average of the explosion epochs derived by \citet{Hamuy2001}, \citet{Leonard2002a,Leonard2002b} and \citet{Elmhamdi2003} using the expanding photosphere method. 
a) this paper;
b) NASA/IPAC Extragalactic Database;   
c) \citealt{1990IAUC4965};
d) \citealt{Schmidt1993};
e) \citealt{Benetti1994};                     
f) \citealt{1999IAUC7294};             
g) \citealt{Hamuy2001};                
h) \citealt{Leonard2002a,Leonard2002b};
i) \citealt{Elmhamdi2003};             
j) \citealt{1999IAUC7329};             
k) \citealt{1999IAUC7334};             
l) \citealt{Leonard2002c};             
m) \citealt{Spiro2014};                
n) \citealt{Poznanski2009};            
o) \citealt{2004IAUC8413};             
p) \citealt{Li2005};                   
q) \citealt{Sahu2006};                 
r) \citealt{Maguire2010b};             
s) \citealt{Pastorello2006};           
t) \citealt{Takats2006};               
u) \citealt{Tsvetkov2006};             
v) \citealt{Brown2007};                
w) \citealt{Dessart2008};              
x) \citealt{Pastorello2009a};          
y) \citealt{Quimby2007};               
z) \citealt{Gal-Yam2011};              
A) \citealt{Tomasella2013};         
B) \citealt{Faran2014b};              
C) \citealt{2001IAUC7655};           
D) \citealt{2013ATel5237};    
E) \citealt{2013CBET3606}; 
F) \citealt{2013CBET3609.4}; 
G) \citealt{Valenti2014b};             
H) \citealt{Richmond2014};             
I) \citealt{Bose2015};
J) \citealt{Balinskaia1980};          
K) \citealt{deVaucouleurs1981};       
L) \citealt{Barbon1982};              
M) \citealt{2001IAUC7737};            
N) \citealt{Elias-Rosa2011};           
O) \citealt{2009CBET1874};             
P) \citealt{2009CBET1867}.             
\end{sidewaystable*}

\subsection{Rise time parametrization}
\label{section:Rise_times}

We investigate the rise times of 20 Type II-P and II-L SNe with explosion epochs constrained between $\pm$ 0.3\,days and $\pm$ 8\,days.

For most SNe the epoch of explosion was adopted from the literature. In the cases where no estimate was given we derived the explosion epoch by taking the average MJD between the first detection and the last pre-discovery non-detection (SNe 1979C, 2000dc, 2001cy, 1999gn, 1999gi, 2009hd, 2001do, 2001fa). For SNe 2004du and 2004et no pre-discovery non-detections were reported, we therefore used a low-order polynomial fit to the observed data to estimate the explosion epoch. Finally for SN 1999em multiple estimates for the explosion epoch using the expanding photosphere method are available. For our analysis we used the mean value dervied from the best estimates of 
\citet{Hamuy2001}, \citet{Leonard2002a,Leonard2002b}, and \citet{Elmhamdi2003}.

In order to derive the rise time in an internally self-consistent manner, we define an epoch
termed ``end-of-rise'' which is the epoch at which the $r'$-band magnitude rises by less than 0.01\,mag.\,d$^{-1}$.
This was estimated by fitting a low-order polynomial to the data, with an iteratively chosen step-size in time.
We found that this approach allowed the inclusion of SNe with different light curve morphologies (Fig.
\ref{figure:LC_rise_comparison}): while the light curves of some SNe (such as LSQ13cuw) have a clear peak and 
a well defined maximum, others (such as SN 1999em) display no clear peak before settling onto the plateau or 
(e.g. SN~2005cs) have a rising $r'$-band light curve after an initial weak maximum, which would lead to an 
epoch of maximum light well into the plateau phase.

To estimate the error in this process, we linearly extrapolated the fit to the point where the gradient equalled zero. The midpoint between this value and either the last measured data point, or the ``end-of-rise'' epoch (whichever one was more distant) was taken to be a conservative estimate of the error. 
This naturally accounts for how well-sampled the rise to maximum or plateau actually was. With the exception of SN 1979C the SNe in our sample have a sampling quality of 0.3-1.3 photometric points per day (averaged over $\pm$ 3\,days around the fitted result for the ``end-of-rise'' epoch), which is reasonably good for our purposes.
The final error on the rise time is given by adding the errors on the explosion epoch and the ``end-of-rise'' epoch in quadrature. Analogously the error on the ``end-of-rise'' absolute magnitude is determined by the errors on the distance modulus, the ``end-of-rise'' magnitude and the total reddening.
The error in the explosion epoch estimate is combined in quadrature with the error in the gradient; the former dominates the combined error in all cases.

\begin{figure}
  \centering
    \includegraphics[width=9cm]{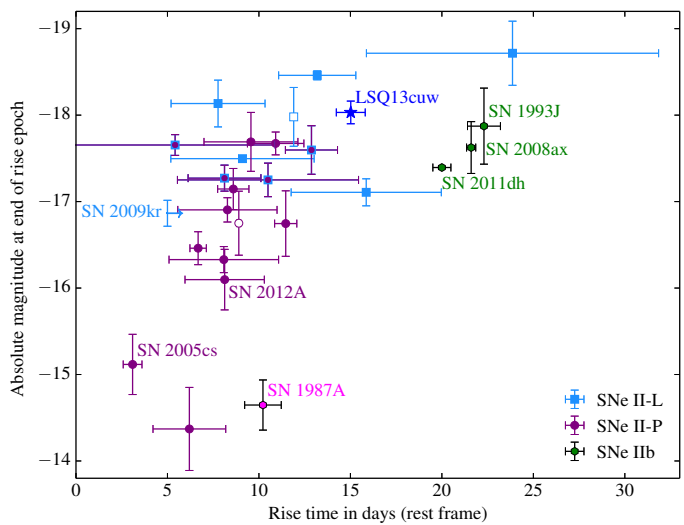}
    \caption{Comparison of ``end-of-rise'' $r'$/$R$-band absolute magnitudes and rise times. The filled squares represent SNe that were classified as Type II-L, while the filled circles are SNe that were classified as Type II-P. The empty square and circle represent rise time measurements for the $R$-band II-L and II-P templates from \citet{Faran2014b}, respectively. As these templates are in a relative scale, we assumed the mean $B$-band values published by \citet{Richardson2014} as the absolute magnitudes for the two SN types, respectively. 
The arrow represents a lower limit for the rise time of SN 2009kr. For the individually labelled SNe 2005cs, 2009kr and 2012A direct progenitor detections have been reported (see also Section \ref{section:progenitor_detections}).
The peculiar Type II-P SN 1987A and the Type IIb SNe 1993J, 2008ax and 2011dh are included for comparison (see Table \ref{table:IIbs}).}
    \label{figure:absmag_vs_rise_nolabels}
\end{figure}

Figure \ref{figure:absmag_vs_rise_nolabels} shows a comparison of the rise times with the absolute magnitudes at the ``end-of-rise'' epoch. 
The filled squares represent SNe that were previously classified as II-L, while the filled circles were classified as SNe II-P. The empty square and circle represent rise time measurements for the II-P and II-L templates from \citet{Faran2014b}, respectively. For the absolute magnitudes of these templates we assumed the mean values published by \citet{Richardson2014} for the two types, respectively. We note, however, that these are $B$-band values. The arrow represents a lower limit for the rise time of the Type II-L SN 2009kr. Unfortunately more than one month passed between the last non-detection and the discovery of SN 2009kr \citep{2009CBET2006} and the explosion epoch is therefore not very well constrained. However, pre-peak photometry for SN 2009kr has been published by \citet{Elias-Rosa2010} and the difference between the discovery and the estimated epoch of ``end-of-rise'' was used as a lower limit for the rise time. We also included the peculiar Type II-P SN 1987A and the Type IIb SNe 1993J, 2008ax, and 2011dh for comparison (see also Table \ref{table:IIbs} in the Appendix). For these SNe the explosion epochs are constrained to less than $\pm$ 1\,d. For the estimate of the rise time the first peak was used in the case of SN 1987A and the second peak was used in the case of SN 1993J. 

\subsection{Rise time correlations}

An inspection of Table \ref{table:SN_all} and Fig. \ref{figure:LC_rise_comparison} reveals some intriguing points which we highlight below. First, only considering those few objects in our sample that have the most tightly constrained explosion epochs ($\lesssim$1\,d) and peak magnitudes, we find that SN~2005cs, an undisputed type II-P SN, rose to $M_R \sim -15$ in $\sim$3\,d, while LSQ13cuw, with type II-L characteristics, had a rise time of $\sim$15\,d and a peak magnitude of $M_r' \sim -18$. Objects with peak magnitudes fainter than LSQ13cuw (such as SNe 2004et, 2006bp) have correspondingly shorter rise times. This is borne out in Fig. \ref{figure:absmag_vs_rise_nolabels} albeit with some scatter.

Other studies have noted the possibility of such a trend. For example, the early discovery of SN~2010id led \citet{Gal-Yam2011} to compare its rise time to that of SNe 2005cs and 2006bp; they found that SNe 2005cs and 2010id -- both faint type II-P SNe -- had shorter rise times than SN~2006bp. On the other hand \citet{Faran2014a} in a recent study, report no conclusive correlation between rise time and peak brightness. 

If the above trend holds over all type II-P, L SNe, then objects such as SNe 2000dc and 2001do appear to be true intermediates in the sense that they are brighter at peak and have longer rise times than the average SN II-P, while they are fainter, and have shorter rise times than the average II-L. Interestingly, these very objects were culled by \citet{Poznanski2009} from their sample of type II-P SNe to be used as distance indicators owing to their larger-than-average post-peak decline rates, while \citet{Faran2014b} classified these as type II-L SNe for their study of type II SNe. 
Another case in point is SN 2013ej: dubbed a ``slow-rising'' type II-P SN by \citet{Valenti2014b}, while \citet{Bose2015} favour a ``Type II-L'' classification based on its post-peak decline rate. We find it to be similar to the ``fast-declining'' SNe II-P in that it is relatively bright, and has a relatively long rise time compared to other SNe II-P (see Table \ref{table:SN_all}).

In Fig. \ref{figure:absmag_vs_rise_nolabels} we show the location of the well-known, but peculiar type II SN~1987A as a reference point. It is faint at peak relative to the other objects, but has a rise time of $\sim$10\,d to the first peak when estimated in a manner identical to the rest of our sample. In the light of our earlier discussion on the possibility of IIb SNe showing linearly declining light curves, we include three well-observed type IIb SNe in Fig. \ref{figure:absmag_vs_rise_nolabels}. With 22.3 $\pm$ 0.9, 21.6 $\pm$ 0.3 and 20.0 $\pm$ 0.5 days the Type IIb SNe 1993J, 2008ax and 2011dh have a significantly longer rise times than the SNe in our sample at a similar brightness ($\sim$ $-$17.4 to $-$17.9\,mag). The only SN with a similar rise time is SN 1979C; however, since its explosion epoch is not very well constrained and its rise time error therefore correspondingly larger ($\pm$ 8\,days). We note that the rise time of SN 1993J, which has a similar luminosity as LSQ13cuw is almost one week longer than that of LSQ13cuw (15.0 $\pm$ 0.8\,days). 

We next consider whether the type II-P SNe rise times are different from those of type II-L SNe. Taken at face value, the weighted average of each subsample seems to indicate that there is a difference: 7.0 $\pm$ 0.3\,d (II-P) versus 13.3 $\pm$ 0.6\,d (II-L). However, we note that the most extreme objects in terms of their rise time duration, also have the most tightly constrained explosion epochs. We might therefore expect that when objects (of either formal subtype) of intermediate peak brightness become available, a similar treatment is likely to result in an overlap between average rise times. We return to this point below. For comparison, the rise time measurements for the II-P and II-L light curves templates constructed from the \citet{Faran2014b} sample using the same method, we find rise times of 8.9 $\pm$ 0.5\,d (II-P) and 11.9 $\pm$ 0.5\,d (II-L). These are shown as open symbols in Figure \ref{figure:absmag_vs_rise_nolabels}. Both samples are small and have objects in common, but there are significant differences in the make up. For instance, the inclusion of SNe known to be interacting with a dense circumstellar medium \citep[e.g. SN~2008fq,][]{Taddia2013}, but which have linearly declining light curves, could arguably be included in studies that consider the post-peak properties of type II-L SNe, but are likely to bias rise time estimates in ways not considered here.

There has been much discussion in the literature regarding the bimodality or otherwise in the properties of type II-P versus type II-L SNe \citep[e.g.][]{Anderson2014a,Sanders2014}. Motivated by our findings above, we suggest that an alternative way of subdividing these two groups of SNe would be to consider their rise time / peak brightness values if available: SNe with longer rise times will generally be brighter at peak. A corollary would be that they would also decline faster post-peak; i.e., these would be the undisputed type II-L SNe in the traditional classification. Currently, constraints on explosion epochs to better than 1\,d are not routinely available. However, this is a technological shortcoming that future planned facilities will remove.

\subsection{Rise time modelling and comparison with observations}
\label{section:Rise_time_modelling}

In this subsection we attempt to put our findings into context with theoretical expectations.
We have restricted ourselves to focussing on the main results here, with full details of the 
models and associated assumptions presented in the Appendix.

The optical light curve can be thought of as the bolometric light curve multiplied by the fraction of the bolometric light that emerges in the optical. As this fraction usually is a faster evolving function than the bolometric luminosity, the optical peak will be close to the peak of the fraction function. We are therefore justified in comparing to $r'$- or $R$-band observations. 

We performed a simple parameter study using the analytical models from \citet{Arnett1980,Arnett1982} and \citet{Rabinak2011} (see Section \ref{section:Analytical_models_for_cooling_phase}) assuming a blackbody spectral energy distribution and no energy input from $^{56}$Ni in the ejecta. The early bolometric light curves were calculated while varying the progenitor radius between 100 and 900\,R$_{\astrosun}$, the explosion energy between $0.1 \times 10^{51}$ and $5 \times 10^{51}$\,erg and the ejecta mass between 3 and 20\,M$_{\astrosun}$. 
The bolometric light curves were also used to derive the evolution in different filters, by artificially setting the object to a distance of 10\,Mpc. Finally the same rise time fitting algorithm was applied to the model light curves as was previously applied to the SN data. 

In comparison, more sophisticated hydrodynamical models allow a proper treatment of the time-dependent radiative transfer equations, however they are very time consuming in their calculations, in particular over a larger parameter space. \citet{Blinnikov1993}, for example, assume explosion energies of E $\sim$ 1-2 $\times 10^{51}$\,erg, ejecta masses of M $\sim$ 1-6\,M$_{\astrosun}$, progenitor radii of R$_0 \sim$ 500-600\,R$_{\astrosun}$ and derive $B$-band peak magnitudes of $\sim -16$\,mag, which is in good agreement with the simple analytic models we use in this paper.

\begin{figure*}[t!]
   \centering
   \begin{subfigure}[t]{0.5\textwidth}
      \includegraphics[width=9cm]{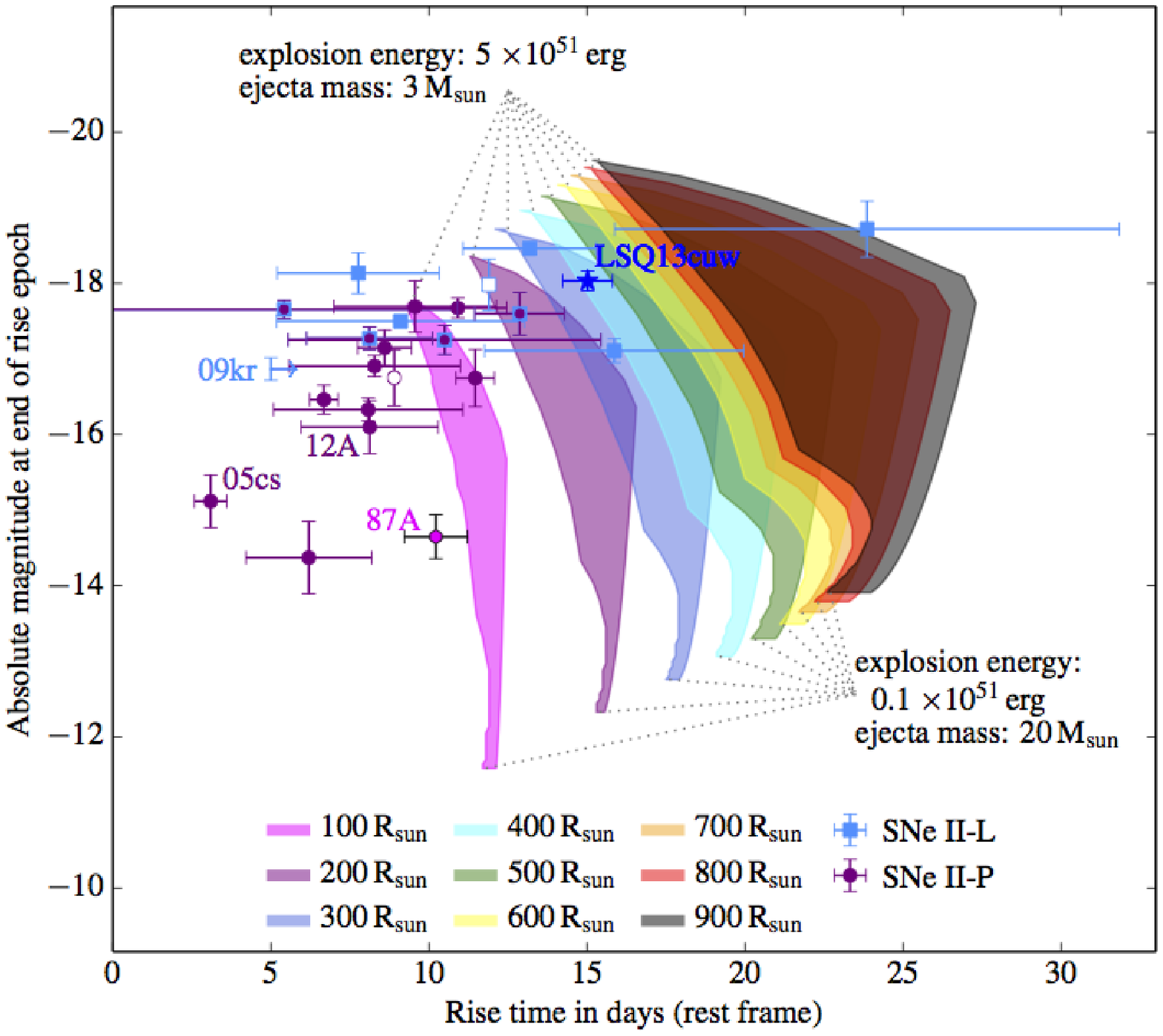}
      \caption{\citet{Arnett1980,Arnett1982} model}
      \label{figure:model_data_risetime_absmag_comparison_Arnett}
   \end{subfigure}%
   \begin{subfigure}[t]{0.5\textwidth}
      \includegraphics[width=9cm]{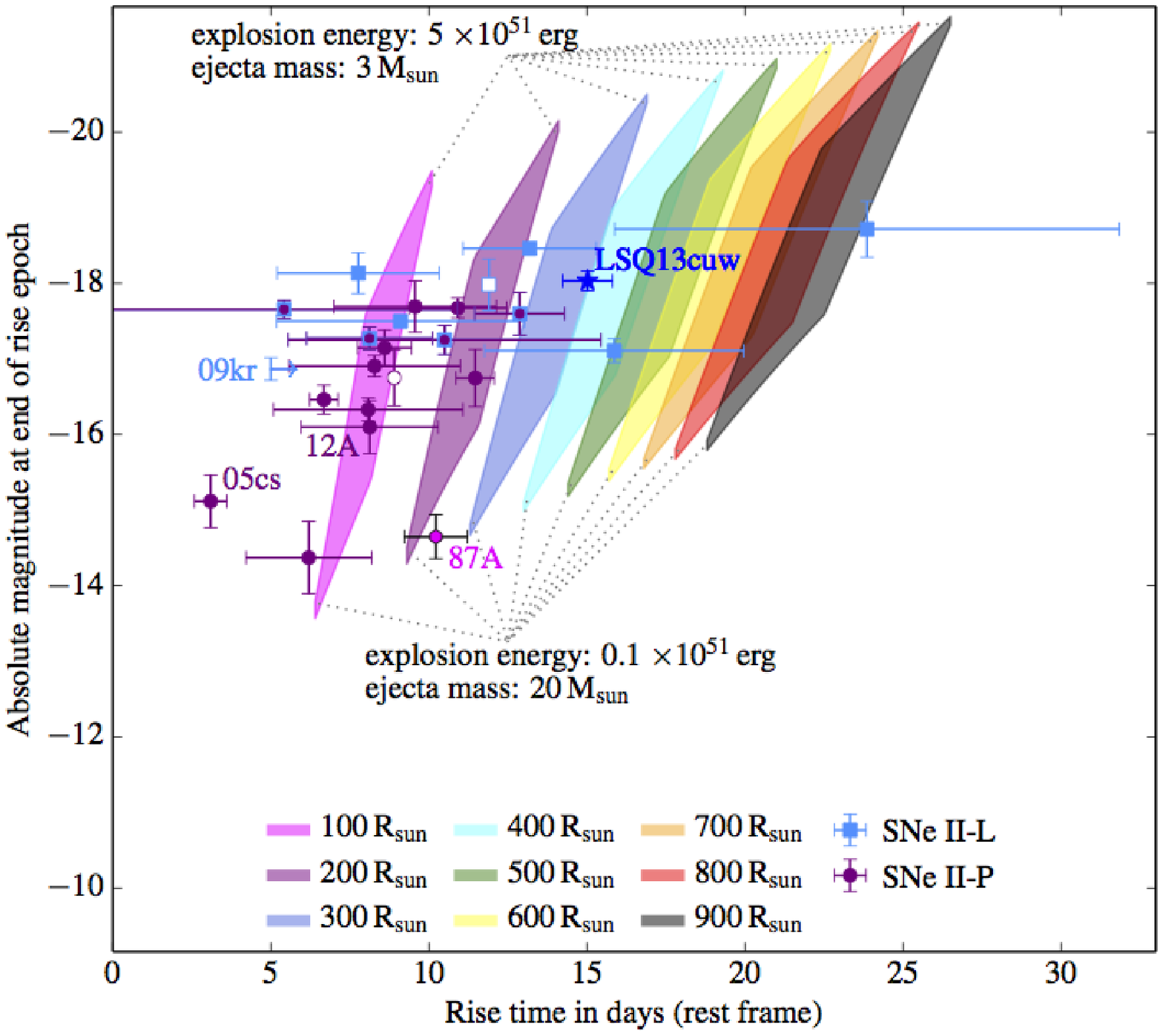}
      \caption{\citet{Rabinak2011} model}
      \label{figure:model_data_risetime_absmag_comparison_Rabinak}
   \end{subfigure}
   \caption{Theoretical rise times and end rise absolute magnitudes using the \citet{Arnett1980,Arnett1982} and \citet{Rabinak2011} model compared with the SNe in our sample. The SNe are depicted in the same way as in Figure \ref{figure:absmag_vs_rise_nolabels}. The shaded regions represent the parameter space for specific progenitor radii, respectively. The input parameters were varied as follows:  100\,R$_{\astrosun} \leq R_0 \leq$ 900\,R$_{\astrosun}$, $0.1 \times 10^{51}$\,erg $\leq E \leq 5 \times 10^{51}$\,erg and 3 $\leq M \leq$ 20\,M$_{\astrosun}$. 
The same rise time fitting algorithm was applied to the model optical light curves as was previously applied to the 
SN data. Only those objects that have progenitor detections are labelled explicitly.
}
   \label{figure:model_data_risetime_absmag_comparison}
\end{figure*}

In Figures \ref{figure:model_data_risetime_absmag_comparison_Arnett} and \ref{figure:model_data_risetime_absmag_comparison_Rabinak} we compare our rise time estimates with the results from the analytic models. 
The SNe are depicted in the same way as in Figure \ref{figure:absmag_vs_rise_nolabels}. The shaded regions represent the parameter space for specific progenitor radii, respectively. For a constant radius the brightest magnitude results from an explosion energy of $5 \times 10^{51}$\,erg and an ejecta mass of 3\,M$_{\astrosun}$, while the faintest magnitude results from an explosion energy of $0.1 \times 10^{51}$\,erg and an ejecta mass of 20\,M$_{\astrosun}$.

The models in general predict somewhat longer rise times than are observed. It is remarkable, however, that even though the models used are simplistic in their physical description, the results from the models lie reasonably close to the measurements from our sample in the magnitude-rise time diagram, and in particular span similar ranges to the data.
The models can therefore give us a relatively accurate picture of the trends involved. 

We found that the model parameters $M$, $E$, and $R_0$ influence the peak luminosity and rise time as follows: 
the luminosity of the optical peak increases with larger $R_0$ and $E$, and decreases with larger $M$ for both the \citet{Arnett1980,Arnett1982} and the \citet{Rabinak2011} model.
For the optical rise time we derived approximate analytical expressions that relate it to the parameters $R_0$, $E$ and $M$ (see Equations \ref{equation:t_optrise_Arnett} and \ref{equation:t_optrise_Rabinak} in the Appendix). For the \citet{Arnett1980,Arnett1982} model we found that $t_{\mathrm{opt-rise}} \propto R_0^{1/2} T_{\mathrm{peak}}^{-2}$, while for the \citet{Rabinak2011} model $t_{\mathrm{opt-rise}} \propto R_0^{0.55} T_{\mathrm{peak}}^{-2.2} E_{51}^{0.06} M^{-0.12}$.
The dependence of the rise time on mass and explosion energy is smaller than the dependence on the radius, as we can see in Equations \ref{equation:t_optrise_Arnett} and \ref{equation:t_optrise_Rabinak}. In fact our estimate for the rise time shows that to first order the effects of mass and energy on the rise time is negligible (in particular for the \citet{Arnett1980,Arnett1982} model).

With these formulae we now have a quantitative handle on the dependency on $R_0$; it is a square root dependency which fits reasonably well with the full solutions (see Figures \ref{figure:model_data_risetime_absmag_comparison_Arnett} and \ref{figure:model_data_risetime_absmag_comparison_Rabinak}).
Using hydrodynamical simulations \citet{Swartz1991} also find that larger progenitor radii result in supernovae with longer rise times and brighter optical peaks.

We have seen in the previous section that the difference between the rise times of SNe II-L and II-P is small, however that SNe II-L tend to have somewhat longer rise times and higher luminosities than SNe II-P. Additionally, from numerical models such as \citet{Blinnikov1993} we expect SNe II-L to have higher $E/M$ values than SNe II-P. In the \citet{Arnett1980,Arnett1982} model a higher $E/M$ gives slightly shorter rise times, so a larger $R_0$ is needed to explain the somewhat longer rise times of SNe II-L. 

The \citet{Rabinak2011} model, on the other hand gives longer rise times for higher $E/M$ values which are of the same order (a few days) as the observed difference between the rise times of SNe II-L and II-P. For the \citet{Rabinak2011} model it is therefore not imminently necessary that the progenitors of SNe II-L have larger radii than the ones of SNe II-P.

\subsection{Rise times of SNe with direct progenitor detections}
\label{section:progenitor_detections}

In the following we consider the cases where both direct detections of a Type II-P/L progenitor and constraints on the rise time are available: the two Type II-P SNe 2005cs and 2012A, the Type II-L SN 2009kr and the peculiar Type II-P SN 1987A. 
A summary of inferred parameters for the progenitors is given in Table \ref{table:progenitor_detections}. 

The parameters can generally be split into two categories: the direct observations of magnitudes and colours which were used in the literature to estimate the mass, the effective temperature, and the luminosity of the progenitor star. We used these in turn to calculate the radius of the progenitor star, assuming it radiates as a blackbody: $L = 4\pi R^2 \sigma T^4$. The second category consists of published models that were tailored to match the observations of each SN; this yields estimates of the progenitor mass, explosion energy, and the progenitor radius. 

For SN 2005cs the estimates for the progenitor radius vary quite significantly, spanning a range from 100 to 700\,R$_{\astrosun}$ (see Table \ref{table:progenitor_detections} and references therein). It is therefore difficult to compare it with the other objects. The two estimates for the progenitor radius of the Type II-L SN 2009kr published by \citet{Elias-Rosa2010} and \citet{Fraser2010} are in agreement with each other, and it seems that the progenitor of SN 2009kr was larger than that of the Type II-P SN 2012A. \citet{Maund2015} however argue that the progenitor previously identified as a yellow supergiant is in fact a compact cluster. 
Nevertheless, the progenitor mass estimates remain within the bounds of previous studies.
SN 2012A has a relatively short rise time of only a few days. The lower limit of the rise time of SN 2009kr is somewhat larger than the rise time of SN 2012A, however not significantly. 

It is interesting to note that \citet{Blinnikov1993} require relatively large radii (600-1000\,R$_{\astrosun}$) to produce SNe II-L with peak $B$-band brightnesses between $-16$ and $-17$\,mag. For the relatively bright SN 1979C they even assume a radius of $R_0$ = 6000\,R$_{\astrosun}$ in their models to match the SN luminosity. 

Even though the results for the progenitor radii are not in one-to-one agreement -- indeed that was not
the intention of our study -- with the values determined values using the \citet{Arnett1982} or the \citet{Rabinak2011} model, the fact that the results are of the same order is encouraging. It is clear that a single analytic model cannot fit both cases like a compact progenitor as for SN 1987A, and extended red giants as for some of the other SNe. However, Figures \ref{figure:model_data_risetime_absmag_comparison_Arnett} and \ref{figure:model_data_risetime_absmag_comparison_Rabinak} show that these simple models succeed in reproducing the parameters of the correct order and can help to explain the observed trends as we have also seen in Section \ref{section:Radius}.

\begin{table*}
  \caption{Direct progenitor detections.}
  \label{table:progenitor_detections}
  \centering
  \begin{tabular}{c c   c               l                              c                     c                       c                    c                                    c                         c                    }
  \hline                                                             
            &          &               &                              & \multicolumn{2}{c}{from modelling}          & \multicolumn{3}{c}{from observations}                                             &                     \\
  SN        & \multicolumn{2}{c}{Type} & $M$ in M$_{\astrosun}$       & $E$                  & $R_0$/R$_{\astrosun}$& $T_{\mathrm{eff}}$ & $L$/L$_{\astrosun}$                & $R_0^*$/R$_{\astrosun}$ &                     \\ 
            & SN       & Prog.         &                              & [10$^{50}$\,erg]     &                      & [K]                &                                    &                         &                     \\ 
\hline                                                               
  SN 1987A  & II-P pec & BSG           & initial: 16-22               &                      &                      & 15000-18000        & $1.04^{+0.52}_{-0.26} \times 10^5$ & 43 $\pm$ 14             & 1                   \\
            &          &               &                              &                      &                      &                    &                                    &                         &                     \\
  SN 2005cs & II-P     & RSG           & initial: $9^{+3}_{-2}$       &                      &                      & 2800-4000          & 1.0-2.8 $\times 10^4$              & 200-700                 & 2                   \\
            &          &               & initial: 10 $\pm$ 3          &                      &                      &                    &                                    &                         & 3                   \\
            &          &               & pre SN: 17.3 $\pm$ 1         & $4.1 \pm 0.3$        & 600 $\pm$ 140        &                    &                                    &                         & 4                   \\
            &          &               & ejected: 8–13                & 3                    & 100                  &                    &                                    &                         & 5                   \\
            &          &               & $^{56}$Ni: 0.0065            & \multirow{2}{*}{1.6} & \multirow{2}{*}{360} &                    &                                    &                         & \multirow{2}{*}{6}  \\
            &          &               & envelope: 9.5                &                      &                      &                    &                                    &                         &                     \\
            &          &               &                              &                      &                      &                    &                                    &                         &                     \\
  SN 2009kr & II-L     & YSG           & initial: 18-24               &                      &                      & 5300 $\pm$ 500     & $10^{5.12 \pm 0.15}$               & 430$^{+190}_{-130}$     & 7                   \\
            &          & YSG           & initial: 15$^{+5}_{-4}$      &                      &                      & 4300-5300          & $10^{5.10 \pm 0.24}$               & 500$^{+330}_{-180}$     & 8                   \\
            &          & compact       & \multirow{2}{*}{initial: 16-25} &                      &                      &                    &                                    &                         & \multirow{2}{*}{9}  \\
            &          & cluster       &                              &                      &                      &                    &                                    &                         &                     \\
            &          &               &                              &                      &                      &                    &                                    &                         &                     \\
  SN 2012A  & II-P     & RSG           & pre SN: $10.5^{+4.5}_{-2}$   & \multirow{2}{*}{4.8} & \multirow{2}{*}{260} &                    &                                    &                         & \multirow{2}{*}{10} \\
            &          &               & ejected: 12.5                &                      &                      &                    &                                    &                         &                     \\
  \hline  
  \end{tabular}
  \\[1.5ex]
  \flushleft
  1) \citet{Arnett1989};  
  2) \citet{Maund2005a};         
  3) \citet{Li2006};             
  4) \citet{Utrobin2008};       
  5) \citet{Pastorello2009a};   
  6) \citet{Spiro2014};
  7) \citet{Elias-Rosa2010};
  8) \citet{Fraser2010};    
  9) \citet{Maund2015};    
  10) \citet{Tomasella2013};
$^*$Radius calculated assuming a blackbody $L = 4\pi R^2 \sigma T^4$    
\end{table*}

The results of \citet{Gonzalez-Gaitan2015}, based on comparisons with the analytical 
models of \citet{Nakar2010} and hydrodynamical models of \citet{Tominaga2009}, are in 
agreement with the above, i.e., that the progenitor radii are too small when compared with 
the expected radii of Type II-P/L SNe. 
On this basis, \citet{Gonzalez-Gaitan2015} conclude that temperatures of red supergiant stars are underestimated, or 
that mixing length prescriptions used in stellar evolution models are to blame. While this may 
indeed be the case, we emphasize that the very useful, but still simplistic framework considered 
both here and in \citet{Gonzalez-Gaitan2015}, serve mainly to indicate relevant trends, and that factors such as the 
explosion energy to ejecta mass ratio will strongly influence the actual rise time.

Interestingly, \citet{Gonzalez-Gaitan2015} find that SNe with slower declining light curves (i.e. SNe II-P-like) have
longer rise times than those with faster declining light cruves (i.e. SNe II-L-like). This is in
direct contrast to our findings. We speculate that this difference might arise as a result of inadequate constraints on the explosion epoch or photometric sampling of the rise to peak 
for the majority of SNe presented in \citet{Gonzalez-Gaitan2015}.

\section{Conclusions}
\label{section:Conclusions}

We presented optical light curves and spectra of the LSQ13cuw, 
a type II-L SN that was discovered unusually early, $\sim$1\,d after explosion.
Deep imaging obtained after the SN had faded away revealed a faint extended source, 
within $0''.5$ of the reported position of SDSS J023957.37-083123.8. We propose that this object is the host galaxy of LSQ13cuw.

In order to maximize the amount of information from our data, we assembled
a sample of 20 II-P/L SNe from the literature with well-constrained explosion epochs.
We see a tendency for SNe that have fainter absolute peak magnitudes ($R$ or $r'$ band)
to have shorter rise times, and vice versa.

We performed a simple parameter study using the analytical models for the light curve cooling phase from \citet{Arnett1980,Arnett1982} and \citet{Rabinak2011}. 
The models in general predict somewhat longer rise times than are observed. It is remarkable, however, that even though the used models are simplistic in their physical description, the results from the models lie in a similar region in the magnitude-rise time diagram to the measurements from our sample. The models can therefore give us a picture of the trends involved. The luminosity of the optical peak increases for larger progenitor radii and explosion energies, and decreases for larger masses for both the \citet{Arnett1980,Arnett1982} and the \citet{Rabinak2011} model.
By deriving a simple analytical expression to estimate the rise time for both models we can show that the dependence of the rise time on mass and explosion energy is smaller than the dependence on the progenitor radius. 

In the \citet{Arnett1980,Arnett1982} model a higher $E/M$ gives slightly shorter rise times, so a larger $R_0$ is needed to explain the somewhat longer rise times of SNe II-L. The \citet{Rabinak2011} model, on the other hand gives longer rise times for higher $E/M$ values which are of the same order (a few days) as the observed difference between the rise times of SNe II-L and II-P. 
For the \citet{Rabinak2011} model it is therefore not imminently necessary that the progenitors of SNe II-L have larger radii than the ones of SNe II-P. Independent of the model, the early optical light curves show that Type II-L SNe have somewhat higher $E/M$ values and/or larger progenitor radii than Type II-P SNe. We can rule out that SNe II-L have significantly smaller progenitor radii than SNe II-P. They are too bright and have too long rise times to fit such a scenario. This is at odds with expectations from single star evolution.

\section*{Acknowledgements} 

We are grateful to T. Faran, A. Filippenko, and D. Poznanski for making the light curve data of SNe 2000dc, 2001cy, 2001do, 2001fa, 2003hf, 2005dq available to us prior to publication, and to N. Elias-Rosa for providing the spectra of SN~2009hd in digital form.

R.K. acknowledges funding from STFC (ST/L000709/1).
S.J.S acknowledges funding from the European Research Council under the European Union’s Seventh Framework Programme (FP7/2007-2013)/ERC Grant agreement number 291222.
S.B. is partially supported by the PRIN-INAF 2014 with the project ``Transient Universe: unveiling new types of stellar explosions with PESSTO''.
M.F. is supported by the European Union FP7 programme through ERC grant number 320360 to G. Gilmore.
S.G. acknowledges support from CONICYT through FONDECYT grant 3130680 and from the Ministry of Economy, Development, and Tourism’s Millennium Science Initiative through grant IC12009, awarded to The Millennium Institute of Astrophysics, MAS.
K.M. is supported by a Marie Curie Intra-European Fellowship, within the 7th European Community Framework Programme (FP7).

This work is based (in part) on observations collected at the European Organisation for Astronomical Research in the Southern Hemisphere, Chile as part of PESSTO, (the Public ESO Spectroscopic Survey for Transient Objects) ESO programme 188.D-3003, 191.D-0935.

This work utilizes data from the 40-inch ESO Schmidt Telescope at the La Silla Observatory in Chile with the large area QUEST camera built at Yale University and Indiana University.

The Liverpool Telescope is operated on the island of La Palma by Liverpool John Moores University in the Spanish Observatorio del Roque de los Muchachos of the Instituto de Astrofisica de Canarias with financial support from the UK Science and Technology Facilities Council.

This research has made use of the NASA/IPAC Extragalactic Database (NED) which is operated by the Jet Propulsion Laboratory, California Institute of Technology, under contract with the National Aeronautics and Space Administration.

Funding for the Sloan Digital Sky Survey (SDSS) has been provided by the Alfred P. Sloan Foundation, the Participating Institutions, the National Aeronautics and Space Administration, the National Science Foundation, the U.S. Department of Energy, the Japanese Monbukagakusho, and the Max Planck Society. The SDSS Web site is http://www.sdss.org/.

The SDSS is managed by the Astrophysical Research Consortium (ARC) for the Participating Institutions. The Participating Institutions are The University of Chicago, Fermilab, the Institute for Advanced Study, the Japan Participation Group, The Johns Hopkins University, Los Alamos National Laboratory, the Max-Planck-Institute for Astronomy (MPIA), the Max-Planck-Institute for Astrophysics (MPA), New Mexico State University, University of Pittsburgh, Princeton University, the United States Naval Observatory, and the University of Washington.

This research has made use of the CfA Supernova Archive, which is funded in part by the National Science Foundation through grant AST 0907903.

\bibliography{bib} \bibliographystyle{aa}

\begin{thebibliography}{148}
\expandafter\ifx\csname natexlab\endcsname\relax\def\natexlab#1{#1}\fi

\bibitem[{{Anderson} {et~al.}(2014{\natexlab{a}}){Anderson}, {Dessart},
  {Gutierrez}, {Hamuy}, {Morrell}, {Phillips}, {Folatelli}, {Stritzinger},
  {Freedman}, {Gonz{\'a}lez-Gait{\'a}n}, {McCarthy}, {Suntzeff}, \&
  {Thomas-Osip}}]{Anderson2014b}
{Anderson}, J.~P., {Dessart}, L., {Gutierrez}, C.~P., {et~al.}
  2014{\natexlab{a}}, \mnras, 441, 671

\bibitem[{{Anderson} {et~al.}(2014{\natexlab{b}}){Anderson},
  {Gonz{\'a}lez-Gait{\'a}n}, {Hamuy}, {Guti{\'e}rrez}, {Stritzinger}, {Olivares
  E.}, {Phillips}, {Schulze}, {Antezana}, {Bolt}, {Campillay}, {Castell{\'o}n},
  {Contreras}, {de Jaeger}, {Folatelli}, {F{\"o}rster}, {Freedman},
  {Gonz{\'a}lez}, {Hsiao}, {Krzemi{\'n}ski}, {Krisciunas}, {Maza}, {McCarthy},
  {Morrell}, {Persson}, {Roth}, {Salgado}, {Suntzeff}, \&
  {Thomas-Osip}}]{Anderson2014a}
{Anderson}, J.~P., {Gonz{\'a}lez-Gait{\'a}n}, S., {Hamuy}, M., {et~al.}
  2014{\natexlab{b}}, \apj, 786, 67

\bibitem[{{Arcavi} {et~al.}(2012){Arcavi}, {Gal-Yam}, {Cenko}, {Fox},
  {Leonard}, {Moon}, {Sand}, {Soderberg}, {Kiewe}, {Yaron}, {Becker}, {Scheps},
  {Birenbaum}, {Chamudot}, \& {Zhou}}]{Arcavi2012}
{Arcavi}, I., {Gal-Yam}, A., {Cenko}, S.~B., {et~al.} 2012, \apjl, 756, L30

\bibitem[{{Arnett}(1980)}]{Arnett1980}
{Arnett}, W.~D. 1980, \apj, 237, 541

\bibitem[{{Arnett}(1982)}]{Arnett1982}
{Arnett}, W.~D. 1982, \apj, 253, 785

\bibitem[{{Arnett} {et~al.}(1989){Arnett}, {Bahcall}, {Kirshner}, \&
  {Woosley}}]{Arnett1989}
{Arnett}, W.~D., {Bahcall}, J.~N., {Kirshner}, R.~P., \& {Woosley}, S.~E. 1989,
  \araa, 27, 629

\bibitem[{{Balinskaia} {et~al.}(1980){Balinskaia}, {Bychkov}, \&
  {Neizvestnyi}}]{Balinskaia1980}
{Balinskaia}, I.~S., {Bychkov}, K.~V., \& {Neizvestnyi}, S.~I. 1980, \aap, 85,
  L19

\bibitem[{{Baltay} {et~al.}(2013){Baltay}, {Rabinowitz}, {Hadjiyska}, {Walker},
  {Nugent}, {Coppi}, {Ellman}, {Feindt}, {McKinnon}, {Horowitz}, \&
  {Effron}}]{Baltay2013}
{Baltay}, C., {Rabinowitz}, D., {Hadjiyska}, E., {et~al.} 2013, \pasp, 125, 683

\bibitem[{{Barbon} {et~al.}(1995){Barbon}, {Benetti}, {Cappellaro}, {Patat},
  {Turatto}, \& {Iijima}}]{Barbon1995}
{Barbon}, R., {Benetti}, S., {Cappellaro}, E., {et~al.} 1995, \aaps, 110, 513

\bibitem[{{Barbon} {et~al.}(1979){Barbon}, {Ciatti}, \& {Rosino}}]{Barbon1979}
{Barbon}, R., {Ciatti}, F., \& {Rosino}, L. 1979, \aap, 72, 287

\bibitem[{{Barbon} {et~al.}(1982){Barbon}, {Ciatti}, {Rosino}, {Ortolani}, \&
  {Rafanelli}}]{Barbon1982}
{Barbon}, R., {Ciatti}, F., {Rosino}, L., {Ortolani}, S., \& {Rafanelli}, P.
  1982, \aap, 116, 43

\bibitem[{{Benetti} {et~al.}(1994){Benetti}, {Cappellaro}, {Turatto}, {della
  Valle}, {Mazzali}, \& {Gouiffes}}]{Benetti1994}
{Benetti}, S., {Cappellaro}, E., {Turatto}, M., {et~al.} 1994, \aap, 285, 147

\bibitem[{{Benetti} {et~al.}(1999){Benetti}, {Turatto}, {Cappellaro},
  {Danziger}, \& {Mazzali}}]{Benetti1999}
{Benetti}, S., {Turatto}, M., {Cappellaro}, E., {Danziger}, I.~J., \&
  {Mazzali}, P.~A. 1999, \mnras, 305, 811

\bibitem[{{Bersier} {et~al.}(2013){Bersier}, {Dennefeld}, {Smartt}, {Smith},
  {Benetti}, {Pastorello}, {Valenti}, {Taubenberger}, {Young}, {Inserra},
  {Sullivan}, {Fraser}, {Gal-Yam}, {Yaron}, {Manulis}, {Scalzo}, {Knapic},
  {Molinaro}, {Smareglia}, {Baltay}, {Ellman}, {Hadjiyska}, {McKinnon},
  {Rabinowitz}, {Walker}, {Feindt}, {Kowalski}, {Nugent}, {Burgett},
  {Chambers}, {Huber}, {Kudritzki}, {Magnier}, {Morgan}, {Stubbs}, {Sweeney},
  {Tonry}, {Waters}, {Draper}, {Metcalfe}, \& {Rest}}]{2013ATel5596}
{Bersier}, D., {Dennefeld}, M., {Smartt}, S.~J., {et~al.} 2013, The
  Astronomer's Telegram, 5596, 1

\bibitem[{{Bersten} {et~al.}(2011){Bersten}, {Benvenuto}, \&
  {Hamuy}}]{Bersten2011}
{Bersten}, M.~C., {Benvenuto}, O., \& {Hamuy}, M. 2011, \apj, 729, 61

\bibitem[{{Blinnikov} {et~al.}(2000){Blinnikov}, {Lundqvist}, {Bartunov},
  {Nomoto}, \& {Iwamoto}}]{Blinnikov2000b}
{Blinnikov}, S., {Lundqvist}, P., {Bartunov}, O., {Nomoto}, K., \& {Iwamoto},
  K. 2000, \apj, 532, 1132

\bibitem[{{Blinnikov} \& {Bartunov}(1993)}]{Blinnikov1993}
{Blinnikov}, S.~I. \& {Bartunov}, O.~S. 1993, \aap, 273, 106

\bibitem[{{Bose} {et~al.}(2015){Bose}, {Sutaria}, {Kumar}, {Duggal}, {Misra},
  {Brown}, {Singh}, {Dwarkadas}, {York}, {Chakraborti}, {Chandola},
  {Dahlstrom}, {Ray}, \& {Safonova}}]{Bose2015}
{Bose}, S., {Sutaria}, F., {Kumar}, B., {et~al.} 2015, ArXiv:1504.06207

\bibitem[{{Branch} {et~al.}(1981){Branch}, {Falk}, {Uomoto}, {Wills}, {McCall},
  \& {Rybski}}]{Branch1981}
{Branch}, D., {Falk}, S.~W., {Uomoto}, A.~K., {et~al.} 1981, \apj, 244, 780

\bibitem[{{Brown} {et~al.}(2007){Brown}, {Dessart}, {Holland}, {Immler},
  {Landsman}, {Blondin}, {Blustin}, {Breeveld}, {Dewangan}, {Gehrels},
  {Hutchins}, {Kirshner}, {Mason}, {Mazzali}, {Milne}, {Modjaz}, \&
  {Roming}}]{Brown2007}
{Brown}, P.~J., {Dessart}, L., {Holland}, S.~T., {et~al.} 2007, \apj, 659, 1488

\bibitem[{{Buta}(1982{\natexlab{a}})}]{Buta1982b}
{Buta}, R.~J. 1982{\natexlab{a}}, \pasp, 94, 744

\bibitem[{{Buta}(1982{\natexlab{b}})}]{Buta1982a}
{Buta}, R.~J. 1982{\natexlab{b}}, \pasp, 94, 578

\bibitem[{{Calzavara} \& {Matzner}(2004)}]{Calzavara2004}
{Calzavara}, A.~J. \& {Matzner}, C.~D. 2004, \mnras, 351, 694

\bibitem[{{Cappellaro} {et~al.}(1995){Cappellaro}, {Danziger}, {della Valle},
  {Gouiffes}, \& {Turatto}}]{Cappellaro1995}
{Cappellaro}, E., {Danziger}, I.~J., {della Valle}, M., {Gouiffes}, C., \&
  {Turatto}, M. 1995, \aap, 293, 723

\bibitem[{{Centurion} {et~al.}(1993){Centurion}, {Vladilo}, {Zurek}, {Robb}, \&
  {Balam}}]{1993IAUC5761}
{Centurion}, M., {Vladilo}, G., {Zurek}, D.~R., {Robb}, R.~M., \& {Balam},
  D.~D. 1993, \iaucirc, 5761, 2

\bibitem[{{Chatzopoulos} {et~al.}(2012){Chatzopoulos}, {Wheeler}, \&
  {Vinko}}]{Chatzopoulos2012}
{Chatzopoulos}, E., {Wheeler}, J.~C., \& {Vinko}, J. 2012, \apj, 746, 121

\bibitem[{{Chugai}(1985)}]{Chugai1985}
{Chugai}, N.~N. 1985, Soviet Astronomy Letters, 11, 148

\bibitem[{{Chugai}(1991)}]{Chugai1991}
{Chugai}, N.~N. 1991, Soviet Astronomy Letters, 17, 210

\bibitem[{{Corwin} {et~al.}(1993){Corwin}, {Porter}, {Neely}, {Cohen},
  {Prugniel}, {Hanzl}, {Mikolajewski}, \& {Wikierski}}]{1993IAUC5742}
{Corwin}, H.~G., {Porter}, A.~C., {Neely}, A.~W., {et~al.} 1993, \iaucirc,
  5742, 1

\bibitem[{{de Vaucouleurs} {et~al.}(1981){de Vaucouleurs}, {de Vaucouleurs},
  {Buta}, {Ables}, \& {Hewitt}}]{deVaucouleurs1981}
{de Vaucouleurs}, G., {de Vaucouleurs}, A., {Buta}, R., {Ables}, H.~D., \&
  {Hewitt}, A.~V. 1981, \pasp, 93, 36

\bibitem[{{Dennefeld} {et~al.}(2013){Dennefeld}, {Bersier}, {Lyman}, {Maund},
  {Pastorello}, {De Cia}, {Benetti}, {Inserra}, {Smartt}, {Smith}, {Young},
  {Sullivan}, {Taubenberger}, {Valenti}, {Fraser}, {Yaron}, {Gal-Yam},
  {Knapic}, {Smareglia}, {Molinaro}, {Baltay}, {Ellman}, {Hadjiyska},
  {McKinnon}, {Rabinowitz}, {Walker}, {Feindt}, {Kowalski}, {Nugent}, \&
  {Wyrzykowski}}]{2013ATel5617}
{Dennefeld}, M., {Bersier}, D., {Lyman}, J., {et~al.} 2013, The Astronomer's
  Telegram, 5617, 1

\bibitem[{{Dessart} {et~al.}(2008){Dessart}, {Blondin}, {Brown}, {Hicken},
  {Hillier}, {Holland}, {Immler}, {Kirshner}, {Milne}, {Modjaz}, \&
  {Roming}}]{Dessart2008}
{Dessart}, L., {Blondin}, S., {Brown}, P.~J., {et~al.} 2008, \apj, 675, 644

\bibitem[{{Dessart} \& {Hillier}(2010)}]{Dessart2010}
{Dessart}, L. \& {Hillier}, D.~J. 2010, \mnras, 405, 2141

\bibitem[{{Dessart} \& {Hillier}(2011)}]{Dessart2011}
{Dessart}, L. \& {Hillier}, D.~J. 2011, \mnras, 410, 1739

\bibitem[{{Dhungana} {et~al.}(2013){Dhungana}, {Vinko}, {Wheeler}, {Silverman},
  {Zheng}, {Kehoe}, {Ferrante}, {Marion}, {Quimby}, {Yuan}, \&
  {Akerlof}}]{2013CBET3609.4}
{Dhungana}, G., {Vinko}, K., {Wheeler}, J.~C., {et~al.} 2013, Central Bureau
  Electronic Telegrams, 3609, 4

\bibitem[{{Drake} {et~al.}(2009){Drake}, {Djorgovski}, {Mahabal}, {Beshore},
  {Larson}, {Graham}, {Williams}, {Christensen}, {Catelan}, {Boattini},
  {Gibbs}, {Hill}, \& {Kowalski}}]{Drake2009}
{Drake}, A.~J., {Djorgovski}, S.~G., {Mahabal}, A., {et~al.} 2009, \apj, 696,
  870

\bibitem[{{Dumont} {et~al.}(1993){Dumont}, {Manda}, {Remis}, {Vandenbroere},
  {Grenon}, {Kulesza}, {Mikolajewski}, {Ruminski}, {Golebiewski}, {Wikierski},
  {Zheng}, {Zhou}, {Wu}, \& {Prugniel}}]{1993IAUC5774}
{Dumont}, M., {Manda}, A., {Remis}, J., {et~al.} 1993, \iaucirc, 5774, 2

\bibitem[{{Eastman} {et~al.}(1994){Eastman}, {Woosley}, {Weaver}, \&
  {Pinto}}]{Eastman1994}
{Eastman}, R.~G., {Woosley}, S.~E., {Weaver}, T.~A., \& {Pinto}, P.~A. 1994,
  \apj, 430, 300

\bibitem[{{Elias-Rosa} {et~al.}(2010){Elias-Rosa}, {Van Dyk}, {Li}, {Miller},
  {Silverman}, {Ganeshalingam}, {Boden}, {Kasliwal}, {Vink{\'o}}, {Cuillandre},
  {Filippenko}, {Steele}, {Bloom}, {Griffith}, {Kleiser}, \&
  {Foley}}]{Elias-Rosa2010}
{Elias-Rosa}, N., {Van Dyk}, S.~D., {Li}, W., {et~al.} 2010, \apjl, 714, L254

\bibitem[{{Elias-Rosa} {et~al.}(2011){Elias-Rosa}, {Van Dyk}, {Li},
  {Silverman}, {Foley}, {Ganeshalingam}, {Mauerhan}, {Kankare}, {Jha},
  {Filippenko}, {Beckman}, {Berger}, {Cuillandre}, \& {Smith}}]{Elias-Rosa2011}
{Elias-Rosa}, N., {Van Dyk}, S.~D., {Li}, W., {et~al.} 2011, \apj, 742, 6

\bibitem[{{Elmhamdi} {et~al.}(2003){Elmhamdi}, {Danziger}, {Chugai},
  {Pastorello}, {Turatto}, {Cappellaro}, {Altavilla}, {Benetti}, {Patat}, \&
  {Salvo}}]{Elmhamdi2003}
{Elmhamdi}, A., {Danziger}, I.~J., {Chugai}, N., {et~al.} 2003, \mnras, 338,
  939

\bibitem[{{Ergon} {et~al.}(2014{\natexlab{a}}){Ergon}, {Jerkstrand},
  {Sollerman}, {Elias-Rosa}, {Fransson}, {Fraser}, {Pastorello}, {Kotak},
  {Taubenberger}, {Tomasella}, {Valenti}, {Benetti}, {Helou}, {Kasliwal},
  {Maund}, {Smartt}, \& {Spyromilio}}]{Ergon2014b}
{Ergon}, M., {Jerkstrand}, A., {Sollerman}, J., {et~al.} 2014{\natexlab{a}},
  ArXiv:1408.0731

\bibitem[{{Ergon} {et~al.}(2014{\natexlab{b}}){Ergon}, {Sollerman}, {Fraser},
  {Pastorello}, {Taubenberger}, {Elias-Rosa}, {Bersten}, {Jerkstrand},
  {Benetti}, {Botticella}, {Fransson}, {Harutyunyan}, {Kotak}, {Smartt},
  {Valenti}, {Bufano}, {Cappellaro}, {Fiaschi}, {Howell}, {Kankare}, {Magill},
  {Mattila}, {Maund}, {Naves}, {Ochner}, {Ruiz}, {Smith}, {Tomasella}, \&
  {Turatto}}]{Ergon2014a}
{Ergon}, M., {Sollerman}, J., {Fraser}, M., {et~al.} 2014{\natexlab{b}}, \aap,
  562, A17

\bibitem[{{Falk}(1978)}]{Falk1978}
{Falk}, S.~W. 1978, \apjl, 225, L133

\bibitem[{{Falk} \& {Arnett}(1977)}]{Falk1977}
{Falk}, S.~W. \& {Arnett}, W.~D. 1977, \apjs, 33, 515

\bibitem[{{Faran} {et~al.}(2014{\natexlab{a}}){Faran}, {Poznanski},
  {Filippenko}, {Chornock}, {Foley}, {Ganeshalingam}, {Leonard}, {Li},
  {Modjaz}, {Nakar}, {Serduke}, \& {Silverman}}]{Faran2014a}
{Faran}, T., {Poznanski}, D., {Filippenko}, A.~V., {et~al.} 2014{\natexlab{a}},
  \mnras, 442, 844

\bibitem[{{Faran} {et~al.}(2014{\natexlab{b}}){Faran}, {Poznanski},
  {Filippenko}, {Chornock}, {Foley}, {Ganeshalingam}, {Leonard}, {Li},
  {Modjaz}, {Serduke}, \& {Silverman}}]{Faran2014b}
{Faran}, T., {Poznanski}, D., {Filippenko}, A.~V., {et~al.} 2014{\natexlab{b}},
  ArXiv:1409.1536

\bibitem[{{Filippenko} \& {Chornock}(2001)}]{2001IAUC7737}
{Filippenko}, A.~V. \& {Chornock}, R. 2001, \iaucirc, 7737, 2

\bibitem[{{Fransson} {et~al.}(2005){Fransson}, {Challis}, {Chevalier},
  {Filippenko}, {Kirshner}, {Kozma}, {Leonard}, {Matheson}, {Baron},
  {Garnavich}, {Jha}, {Leibundgut}, {Lundqvist}, {Pun}, {Wang}, \&
  {Wheeler}}]{Fransson2005}
{Fransson}, C., {Challis}, P.~M., {Chevalier}, R.~A., {et~al.} 2005, \apj, 622,
  991

\bibitem[{{Fraser} {et~al.}(2010){Fraser}, {Tak{\'a}ts}, {Pastorello},
  {Smartt}, {Mattila}, {Botticella}, {Valenti}, {Ergon}, {Sollerman}, {Arcavi},
  {Benetti}, {Bufano}, {Crockett}, {Danziger}, {Gal-Yam}, {Maund},
  {Taubenberger}, \& {Turatto}}]{Fraser2010}
{Fraser}, M., {Tak{\'a}ts}, K., {Pastorello}, A., {et~al.} 2010, \apjl, 714,
  L280

\bibitem[{{Gal-Yam} {et~al.}(2011){Gal-Yam}, {Kasliwal}, {Arcavi}, {Green},
  {Yaron}, {Ben-Ami}, {Xu}, {Sternberg}, {Quimby}, {Kulkarni}, {Ofek},
  {Walters}, {Nugent}, {Poznanski}, {Bloom}, {Cenko}, {Filippenko}, {Li},
  {Silverman}, {Walker}, {Sullivan}, {Maguire}, {Howell}, {Mazzali}, {Frail},
  {Bersier}, {James}, {Akerlof}, {Yuan}, {Law}, {Fox}, \&
  {Gehrels}}]{Gal-Yam2011}
{Gal-Yam}, A., {Kasliwal}, M.~M., {Arcavi}, I., {et~al.} 2011, \apj, 736, 159

\bibitem[{{Ganeshalingam} {et~al.}(2001){Ganeshalingam}, {Modjaz}, \&
  {Li}}]{2001IAUC7655}
{Ganeshalingam}, M., {Modjaz}, M., \& {Li}, W.~D. 2001, \iaucirc, 7655, 1

\bibitem[{{Gezari} {et~al.}(2008){Gezari}, {Dessart}, {Basa}, {Martin},
  {Neill}, {Woosley}, {Hillier}, {Bazin}, {Forster}, {Friedman}, {Le Du},
  {Mazure}, {Morrissey}, {Neff}, {Schiminovich}, \& {Wyder}}]{Gezari2008}
{Gezari}, S., {Dessart}, L., {Basa}, S., {et~al.} 2008, \apjl, 683, L131

\bibitem[{{Gonzalez-Gaitan} {et~al.}(2015){Gonzalez-Gaitan}, {Tominaga},
  {Molina}, {Galbany}, {Bufano}, {Anderson}, {Gutierrez}, {Forster}, {Pignata},
  {Bersten}, {Howell}, {Sullivan}, {Carlberg}, {de Jaeger}, {Hamuy},
  {Baklanov}, \& {Blinnikov}}]{Gonzalez-Gaitan2015}
{Gonzalez-Gaitan}, S., {Tominaga}, N., {Molina}, J., {et~al.} 2015,
  ArXiv:1505.02988

\bibitem[{{Grassberg} {et~al.}(1971){Grassberg}, {Imshennik}, \&
  {Nadyozhin}}]{Grassberg1971}
{Grassberg}, E.~K., {Imshennik}, V.~S., \& {Nadyozhin}, D.~K. 1971, \apss, 10,
  28

\bibitem[{{Grefenstette} {et~al.}(2014){Grefenstette}, {Harrison}, {Boggs},
  {Reynolds}, {Fryer}, {Madsen}, {Wik}, {Zoglauer}, {Ellinger}, {Alexander},
  {An}, {Barret}, {Christensen}, {Craig}, {Forster}, {Giommi}, {Hailey},
  {Hornstrup}, {Kaspi}, {Kitaguchi}, {Koglin}, {Mao}, {Miyasaka}, {Mori},
  {Perri}, {Pivovaroff}, {Puccetti}, {Rana}, {Stern}, {Westergaard}, \&
  {Zhang}}]{Grefenstette2014}
{Grefenstette}, B.~W., {Harrison}, F.~A., {Boggs}, S.~E., {et~al.} 2014, \nat,
  506, 339

\bibitem[{{Guti{\'e}rrez} {et~al.}(2014){Guti{\'e}rrez}, {Anderson}, {Hamuy},
  {Gonz{\'a}lez-Gait{\'a}n}, {Folatelli}, {Morrell}, {Stritzinger}, {Phillips},
  {McCarthy}, {Suntzeff}, \& {Thomas-Osip}}]{Gutierrez2014}
{Guti{\'e}rrez}, C.~P., {Anderson}, J.~P., {Hamuy}, M., {et~al.} 2014, \apjl,
  786, L15

\bibitem[{{Hamuy} {et~al.}(2001){Hamuy}, {Pinto}, {Maza}, {Suntzeff},
  {Phillips}, {Eastman}, {Smith}, {Corbally}, {Burstein}, {Li}, {Ivanov},
  {Moro-Martin}, {Strolger}, {de Souza}, {dos Anjos}, {Green}, {Pickering},
  {Gonz{\'a}lez}, {Antezana}, {Wischnjewsky}, {Galaz}, {Roth}, {Persson}, \&
  {Schommer}}]{Hamuy2001}
{Hamuy}, M., {Pinto}, P.~A., {Maza}, J., {et~al.} 2001, \apj, 558, 615

\bibitem[{{Hamuy} {et~al.}(1988){Hamuy}, {Suntzeff}, {Gonzalez}, \&
  {Martin}}]{Hamuy1988}
{Hamuy}, M., {Suntzeff}, N.~B., {Gonzalez}, R., \& {Martin}, G. 1988, \aj, 95,
  63

\bibitem[{{Hanzl} {et~al.}(1993){Hanzl}, {Kulesza}, {Mikolajewski}, {Ruminski},
  {Staniewski}, \& {Prugniel}}]{1993IAUC5765}
{Hanzl}, D., {Kulesza}, B., {Mikolajewski}, M., {et~al.} 1993, \iaucirc, 5765,
  2

\bibitem[{{Herant} \& {Benz}(1991)}]{Herant1991}
{Herant}, M. \& {Benz}, W. 1991, \apjl, 370, L81

\bibitem[{{Hillebrandt} \& {M{\"u}ller}(1981)}]{Hillebrandt1981}
{Hillebrandt}, W. \& {M{\"u}ller}, E. 1981, \aap, 103, 147

\bibitem[{{Holwerda} {et~al.}(2015){Holwerda}, {Reynolds}, {Smith}, \&
  {Kraan-Korteweg}}]{Holwerda2015}
{Holwerda}, B.~W., {Reynolds}, A., {Smith}, M., \& {Kraan-Korteweg}, R.~C.
  2015, \mnras, 446, 3768

\bibitem[{{Jeffery} \& {Branch}(1990)}]{Jeffery1990}
{Jeffery}, D.~J. \& {Branch}, D. 1990, in Supernovae, Jerusalem Winter School
  for Theoretical Physics, ed. J.~C. {Wheeler}, T.~{Piran}, \& S.~{Weinberg},
  149

\bibitem[{{Jeffery} {et~al.}(1994){Jeffery}, {Kirshner}, {Challis}, {Pun},
  {Filippenko}, {Matheson}, {Branch}, {Chevalier}, {Fransson}, {Panagia},
  {Wagoner}, {Wheeler}, \& {Clocchiatti}}]{Jeffery1994}
{Jeffery}, D.~J., {Kirshner}, R.~P., {Challis}, P.~M., {et~al.} 1994, \apjl,
  421, L27

\bibitem[{{Karp} {et~al.}(1977){Karp}, {Lasher}, {Chan}, \&
  {Salpeter}}]{Karp1977}
{Karp}, A.~H., {Lasher}, G., {Chan}, K.~L., \& {Salpeter}, E.~E. 1977, \apj,
  214, 161

\bibitem[{{Kasen} \& {Woosley}(2009)}]{Kasen2009b}
{Kasen}, D. \& {Woosley}, S.~E. 2009, \apj, 703, 2205

\bibitem[{{Kasliwal} {et~al.}(2009){Kasliwal}, {Sahu}, \&
  {Anupama}}]{2009CBET1874}
{Kasliwal}, M.~M., {Sahu}, D.~K., \& {Anupama}, G.~C. 2009, Central Bureau
  Electronic Telegrams, 1874, 2

\bibitem[{{Kim} {et~al.}(2013){Kim}, {Zheng}, {Li}, {Filippenko}, {Cenko},
  {Richmond}, {Amorim}, {Balam}, {Graham}, \& {Hsiao}}]{2013CBET3606}
{Kim}, M., {Zheng}, W., {Li}, W., {et~al.} 2013, Central Bureau Electronic
  Telegrams, 3606, 1

\bibitem[{{Kj{\ae}r} {et~al.}(2010){Kj{\ae}r}, {Leibundgut}, {Fransson},
  {Jerkstrand}, \& {Spyromilio}}]{Kjaer2010}
{Kj{\ae}r}, K., {Leibundgut}, B., {Fransson}, C., {Jerkstrand}, A., \&
  {Spyromilio}, J. 2010, \aap, 517, A51

\bibitem[{{Klein} \& {Chevalier}(1978)}]{Klein1978}
{Klein}, R.~I. \& {Chevalier}, R.~A. 1978, \apjl, 223, L109

\bibitem[{{Kotak} {et~al.}(2009){Kotak}, {Meikle}, {Farrah}, {Gerardy},
  {Foley}, {Van Dyk}, {Fransson}, {Lundqvist}, {Sollerman}, {Fesen},
  {Filippenko}, {Mattila}, {Silverman}, {Andersen}, {H{\"o}flich}, {Pozzo}, \&
  {Wheeler}}]{Kotak2009}
{Kotak}, R., {Meikle}, W.~P.~S., {Farrah}, D., {et~al.} 2009, \apj, 704, 306

\bibitem[{{Landolt}(1992)}]{Landolt1992a}
{Landolt}, A.~U. 1992, \aj, 104, 340

\bibitem[{{Larsson} {et~al.}(2013){Larsson}, {Fransson}, {Kjaer}, {Jerkstrand},
  {Kirshner}, {Leibundgut}, {Lundqvist}, {Mattila}, {McCray}, {Sollerman},
  {Spyromilio}, \& {Wheeler}}]{Larsson2013}
{Larsson}, J., {Fransson}, C., {Kjaer}, K., {et~al.} 2013, \apj, 768, 89

\bibitem[{{Leonard}(2002)}]{Leonard2002b}
{Leonard}, D.~C. 2002, \pasp, 114, 1291

\bibitem[{{Leonard} {et~al.}(2001){Leonard}, {Filippenko}, {Ardila}, \&
  {Brotherton}}]{Leonard2001}
{Leonard}, D.~C., {Filippenko}, A.~V., {Ardila}, D.~R., \& {Brotherton}, M.~S.
  2001, \apj, 553, 861

\bibitem[{{Leonard} {et~al.}(2006){Leonard}, {Filippenko}, {Ganeshalingam},
  {Serduke}, {Li}, {Swift}, {Gal-Yam}, {Foley}, {Fox}, {Park}, {Hoffman}, \&
  {Wong}}]{Leonard2006}
{Leonard}, D.~C., {Filippenko}, A.~V., {Ganeshalingam}, M., {et~al.} 2006,
  \nat, 440, 505

\bibitem[{{Leonard} {et~al.}(2002{\natexlab{a}}){Leonard}, {Filippenko},
  {Gates}, {Li}, {Eastman}, {Barth}, {Bus}, {Chornock}, {Coil}, {Frink},
  {Grady}, {Harris}, {Malkan}, {Matheson}, {Quirrenbach}, \&
  {Treffers}}]{Leonard2002a}
{Leonard}, D.~C., {Filippenko}, A.~V., {Gates}, E.~L., {et~al.}
  2002{\natexlab{a}}, \pasp, 114, 35

\bibitem[{{Leonard} {et~al.}(2002{\natexlab{b}}){Leonard}, {Filippenko}, {Li},
  {Matheson}, {Kirshner}, {Chornock}, {Van Dyk}, {Berlind}, {Calkins},
  {Challis}, {Garnavich}, {Jha}, \& {Mahdavi}}]{Leonard2002c}
{Leonard}, D.~C., {Filippenko}, A.~V., {Li}, W., {et~al.} 2002{\natexlab{b}},
  \aj, 124, 2490

\bibitem[{{Lewis} {et~al.}(1994){Lewis}, {Walton}, {Meikle}, {Martin},
  {Cumming}, {Catchpole}, {Arevalo}, {Argyle}, {Benn}, {Bunclark}, {Castaneda},
  {Centurion}, {Clegg}, {Delgado}, {Dhillon}, {Goudfrooij}, {Harlaftis},
  {Hassall}, {Helmer}, {Hill}, {Jones}, {King}, {Lazaro}, {Lucey}, {Martin},
  {Miller}, {Morrison}, {Penny}, {Perez}, {Read}, {Rudd}, {Rutten}, {Sharples},
  {Unger}, \& {Vilchez}}]{Lewis1994}
{Lewis}, J.~R., {Walton}, N.~A., {Meikle}, W.~P.~S., {et~al.} 1994, \mnras,
  266, L27

\bibitem[{{Li} {et~al.}(2011){Li}, {Leaman}, {Chornock}, {Filippenko},
  {Poznanski}, {Ganeshalingam}, {Wang}, {Modjaz}, {Jha}, {Foley}, \&
  {Smith}}]{Li2011a}
{Li}, W., {Leaman}, J., {Chornock}, R., {et~al.} 2011, \mnras, 412, 1441

\bibitem[{{Li} {et~al.}(2005){Li}, {Van Dyk}, {Filippenko}, \&
  {Cuillandre}}]{Li2005}
{Li}, W., {Van Dyk}, S.~D., {Filippenko}, A.~V., \& {Cuillandre}, J.-C. 2005,
  \pasp, 117, 121

\bibitem[{{Li} {et~al.}(2006){Li}, {Van Dyk}, {Filippenko}, {Cuillandre},
  {Jha}, {Bloom}, {Riess}, \& {Livio}}]{Li2006}
{Li}, W., {Van Dyk}, S.~D., {Filippenko}, A.~V., {et~al.} 2006, \apj, 641, 1060

\bibitem[{{Li}(1999)}]{1999IAUC7294}
{Li}, W.~D. 1999, \iaucirc, 7294, 1

\bibitem[{{Litvinova} \& {Nadezhin}(1983)}]{Litvinova1983}
{Litvinova}, I.~I. \& {Nadezhin}, D.~K. 1983, \apss, 89, 89

\bibitem[{{Litvinova} \& {Nadezhin}(1985)}]{Litvinova1985}
{Litvinova}, I.~Y. \& {Nadezhin}, D.~K. 1985, Soviet Astronomy Letters, 11, 145

\bibitem[{{Maguire} {et~al.}(2010){Maguire}, {Di Carlo}, {Smartt},
  {Pastorello}, {Tsvetkov}, {Benetti}, {Spiro}, {Arkharov}, {Beccari},
  {Botticella}, {Cappellaro}, {Cristallo}, {Dolci}, {Elias-Rosa}, {Fiaschi},
  {Gorshanov}, {Harutyunyan}, {Larionov}, {Navasardyan}, {Pietrinferni},
  {Raimondo}, {di Rico}, {Valenti}, {Valentini}, \& {Zampieri}}]{Maguire2010b}
{Maguire}, K., {Di Carlo}, E., {Smartt}, S.~J., {et~al.} 2010, \mnras, 404, 981

\bibitem[{{Matheson} {et~al.}(2000){Matheson}, {Filippenko}, {Barth}, {Ho},
  {Leonard}, {Bershady}, {Davis}, {Finley}, {Fisher}, {Gonz{\'a}lez}, {Hawley},
  {Koo}, {Li}, {Lonsdale}, {Schlegel}, {Smith}, {Spinrad}, \&
  {Wirth}}]{Matheson2000a}
{Matheson}, T., {Filippenko}, A.~V., {Barth}, A.~J., {et~al.} 2000, \aj, 120,
  1487

\bibitem[{{Matzner} \& {McKee}(1999)}]{Matzner1999}
{Matzner}, C.~D. \& {McKee}, C.~F. 1999, \apj, 510, 379

\bibitem[{{Maund} {et~al.}(2015){Maund}, {Fraser}, {Reilly}, {Ergon}, \&
  {Mattila}}]{Maund2015}
{Maund}, J.~R., {Fraser}, M., {Reilly}, E., {Ergon}, M., \& {Mattila}, S. 2015,
  \mnras, 447, 3207

\bibitem[{{Maund} {et~al.}(2005){Maund}, {Smartt}, \& {Danziger}}]{Maund2005a}
{Maund}, J.~R., {Smartt}, S.~J., \& {Danziger}, I.~J. 2005, \mnras, 364, L33

\bibitem[{{Monard}(2009)}]{2009CBET1867}
{Monard}, L.~A.~G. 2009, Central Bureau Electronic Telegrams, 1867, 1

\bibitem[{{M{\"u}ller} {et~al.}(1991){M{\"u}ller}, {Fryxell}, \&
  {Arnett}}]{Mueller1991}
{M{\"u}ller}, E., {Fryxell}, B., \& {Arnett}, D. 1991, \aap, 251, 505

\bibitem[{{Nagy} {et~al.}(2014){Nagy}, {Ordasi}, {Vink{\'o}}, \&
  {Wheeler}}]{Nagy2014}
{Nagy}, A.~P., {Ordasi}, A., {Vink{\'o}}, J., \& {Wheeler}, J.~C. 2014, \aap,
  571, A77

\bibitem[{{Nakano} \& {Kushida}(1999)}]{1999IAUC7329}
{Nakano}, S. \& {Kushida}, R. 1999, \iaucirc, 7329, 1

\bibitem[{{Nakano} {et~al.}(2009){Nakano}, {Yusa}, \& {Kadota}}]{2009CBET2006}
{Nakano}, S., {Yusa}, T., \& {Kadota}, K. 2009, Central Bureau Electronic
  Telegrams, 2006, 1

\bibitem[{{Nakar} \& {Sari}(2010)}]{Nakar2010}
{Nakar}, E. \& {Sari}, R. 2010, \apj, 725, 904

\bibitem[{{Nomoto} {et~al.}(1995){Nomoto}, {Iwamoto}, \& {Suzuki}}]{Nomoto1995}
{Nomoto}, K.~I., {Iwamoto}, K., \& {Suzuki}, T. 1995, \physrep, 256, 173

\bibitem[{{Nugent} {et~al.}(2011){Nugent}, {Sullivan}, {Cenko}, {Thomas},
  {Kasen}, {Howell}, {Bersier}, {Bloom}, {Kulkarni}, {Kandrashoff},
  {Filippenko}, {Silverman}, {Marcy}, {Howard}, {Isaacson}, {Maguire},
  {Suzuki}, {Tarlton}, {Pan}, {Bildsten}, {Fulton}, {Parrent}, {Sand},
  {Podsiadlowski}, {Bianco}, {Dilday}, {Graham}, {Lyman}, {James}, {Kasliwal},
  {Law}, {Quimby}, {Hook}, {Walker}, {Mazzali}, {Pian}, {Ofek}, {Gal-Yam}, \&
  {Poznanski}}]{Nugent2011}
{Nugent}, P.~E., {Sullivan}, M., {Cenko}, S.~B., {et~al.} 2011, \nat, 480, 344

\bibitem[{{Panagia} {et~al.}(1980){Panagia}, {Vettolani}, {Boksenberg},
  {Ciatti}, {Ortolani}, {Rafanelli}, {Rosino}, {Gordon}, {Reimers}, {Hempe},
  {Benvenuti}, {Clavel}, {Heck}, {Penston}, {Macchetto}, {Stickland},
  {Bergeron}, {Tarenghi}, {Marano}, {Palumbo}, {Parmar}, {Pollard}, {Sanford},
  {Sargent}, {Sramek}, {Weiler}, \& {Matzik}}]{Panagia1980}
{Panagia}, N., {Vettolani}, G., {Boksenberg}, A., {et~al.} 1980, \mnras, 192,
  861

\bibitem[{{Pastorello} {et~al.}(2008){Pastorello}, {Kasliwal}, {Crockett},
  {Valenti}, {Arbour}, {Itagaki}, {Kaspi}, {Gal-Yam}, {Smartt}, {Griffith},
  {Maguire}, {Ofek}, {Seymour}, {Stern}, \& {Wiethoff}}]{Pastorello2008}
{Pastorello}, A., {Kasliwal}, M.~M., {Crockett}, R.~M., {et~al.} 2008, \mnras,
  389, 955

\bibitem[{{Pastorello} {et~al.}(2006){Pastorello}, {Sauer}, {Taubenberger},
  {Mazzali}, {Nomoto}, {Kawabata}, {Benetti}, {Elias-Rosa}, {Harutyunyan},
  {Navasardyan}, {Zampieri}, {Iijima}, {Botticella}, {di Rico}, {Del Principe},
  {Dolci}, {Gagliardi}, {Ragni}, \& {Valentini}}]{Pastorello2006}
{Pastorello}, A., {Sauer}, D., {Taubenberger}, S., {et~al.} 2006, \mnras, 370,
  1752

\bibitem[{{Pastorello} {et~al.}(2009){Pastorello}, {Valenti}, {Zampieri},
  {Navasardyan}, {Taubenberger}, {Smartt}, {Arkharov}, {B{\"a}rnbantner},
  {Barwig}, {Benetti}, {Birtwhistle}, {Botticella}, {Cappellaro}, {Del
  Principe}, {di Mille}, {di Rico}, {Dolci}, {Elias-Rosa}, {Efimova},
  {Fiedler}, {Harutyunyan}, {H{\"o}flich}, {Kloehr}, {Larionov}, {Lorenzi},
  {Maund}, {Napoleone}, {Ragni}, {Richmond}, {Ries}, {Spiro}, {Temporin},
  {Turatto}, \& {Wheeler}}]{Pastorello2009a}
{Pastorello}, A., {Valenti}, S., {Zampieri}, L., {et~al.} 2009, \mnras, 394,
  2266

\bibitem[{{Patat} {et~al.}(1994){Patat}, {Barbon}, {Cappellaro}, \&
  {Turatto}}]{Patat1994}
{Patat}, F., {Barbon}, R., {Cappellaro}, E., \& {Turatto}, M. 1994, \aap, 282,
  731

\bibitem[{{Pennypacker} {et~al.}(1990){Pennypacker}, {Perlmutter}, {Shara},
  {Sargent}, {Stathakis}, \& {Cannon}}]{1990IAUC4965}
{Pennypacker}, C., {Perlmutter}, S., {Shara}, M., {et~al.} 1990, \iaucirc,
  4965, 1

\bibitem[{{Phillips} {et~al.}(2013){Phillips}, {Simon}, {Morrell}, {Burns},
  {Cox}, {Foley}, {Karakas}, {Patat}, {Sternberg}, {Williams}, {Gal-Yam},
  {Hsiao}, {Leonard}, {Persson}, {Stritzinger}, {Thompson}, {Campillay},
  {Contreras}, {Folatelli}, {Freedman}, {Hamuy}, {Roth}, {Shields}, {Suntzeff},
  {Chomiuk}, {Ivans}, {Madore}, {Penprase}, {Perley}, {Pignata}, {Preston}, \&
  {Soderberg}}]{Phillips2013}
{Phillips}, M.~M., {Simon}, J.~D., {Morrell}, N., {et~al.} 2013, \apj, 779, 38

\bibitem[{{Popov}(1993)}]{Popov1993}
{Popov}, D.~V. 1993, \apj, 414, 712

\bibitem[{{Poznanski} {et~al.}(2009){Poznanski}, {Butler}, {Filippenko},
  {Ganeshalingam}, {Li}, {Bloom}, {Chornock}, {Foley}, {Nugent}, {Silverman},
  {Cenko}, {Gates}, {Leonard}, {Miller}, {Modjaz}, {Serduke}, {Smith}, {Swift},
  \& {Wong}}]{Poznanski2009}
{Poznanski}, D., {Butler}, N., {Filippenko}, A.~V., {et~al.} 2009, \apj, 694,
  1067

\bibitem[{{Poznanski} {et~al.}(2012){Poznanski}, {Prochaska}, \&
  {Bloom}}]{Poznanski2012}
{Poznanski}, D., {Prochaska}, J.~X., \& {Bloom}, J.~S. 2012, \mnras, 426, 1465

\bibitem[{{Pressberger} {et~al.}(1993){Pressberger}, {Maitzen}, \&
  {Neely}}]{1993IAUC5832}
{Pressberger}, R., {Maitzen}, H.~M., \& {Neely}, A.~W. 1993, \iaucirc, 5832, 2

\bibitem[{{Pumo} \& {Zampieri}(2011)}]{Pumo2011}
{Pumo}, M.~L. \& {Zampieri}, L. 2011, \apj, 741, 41

\bibitem[{{Quimby} {et~al.}(2007){Quimby}, {Wheeler}, {H{\"o}flich}, {Akerlof},
  {Brown}, \& {Rykoff}}]{Quimby2007}
{Quimby}, R.~M., {Wheeler}, J.~C., {H{\"o}flich}, P., {et~al.} 2007, \apj, 666,
  1093

\bibitem[{{Rabinak} \& {Waxman}(2011)}]{Rabinak2011}
{Rabinak}, I. \& {Waxman}, E. 2011, \apj, 728, 63

\bibitem[{{Richardson} {et~al.}(2014){Richardson}, {Jenkins}, {Wright}, \&
  {Maddox}}]{Richardson2014}
{Richardson}, D., {Jenkins}, III, R.~L., {Wright}, J., \& {Maddox}, L. 2014,
  \aj, 147, 118

\bibitem[{{Richmond}(2014)}]{Richmond2014}
{Richmond}, M.~W. 2014, Journal of the American Association of Variable Star
  Observers (JAAVSO), 42, 333

\bibitem[{{Ripero} {et~al.}(1993){Ripero}, {Garcia}, {Rodriguez}, {Pujol},
  {Filippenko}, {Treffers}, {Paik}, {Davis}, {Schlegel}, {Hartwick}, {Balam},
  {Zurek}, {Robb}, {Garnavich}, \& {Hong}}]{1993IAUC5731}
{Ripero}, J., {Garcia}, F., {Rodriguez}, D., {et~al.} 1993, \iaucirc, 5731, 1

\bibitem[{{Rogers} \& {Iglesias}(1992)}]{Rogers1992}
{Rogers}, F.~J. \& {Iglesias}, C.~A. 1992, \apjs, 79, 507

\bibitem[{{Sahu} {et~al.}(2013){Sahu}, {Anupama}, \& {Chakradhari}}]{Sahu2013}
{Sahu}, D.~K., {Anupama}, G.~C., \& {Chakradhari}, N.~K. 2013, \mnras, 433, 2

\bibitem[{{Sahu} {et~al.}(2006){Sahu}, {Anupama}, {Srividya}, \&
  {Muneer}}]{Sahu2006}
{Sahu}, D.~K., {Anupama}, G.~C., {Srividya}, S., \& {Muneer}, S. 2006, \mnras,
  372, 1315

\bibitem[{{Sanders} {et~al.}(2014){Sanders}, {Soderberg}, {Gezari},
  {Betancourt}, {Chornock}, {Berger}, {Foley}, {Challis}, {Drout}, {Kirshner},
  {Lunnan}, {Marion}, {Margutti}, {McKinnon}, {Milisavljevic}, {Narayan},
  {Rest}, {Kankare}, {Mattila}, {Smartt}, {Huber}, {Burgett}, {Draper},
  {Hodapp}, {Kaiser}, {Kudritzki}, {Magnier}, {Metcalfe}, {Morgan}, {Price},
  {Tonry}, {Wainscoat}, \& {Waters}}]{Sanders2014}
{Sanders}, N.~E., {Soderberg}, A.~M., {Gezari}, S., {et~al.} 2014,
  ArXiv:1404.2004

\bibitem[{{Schawinski} {et~al.}(2008){Schawinski}, {Justham}, {Wolf},
  {Podsiadlowski}, {Sullivan}, {Steenbrugge}, {Bell}, {R{\"o}ser}, {Walker},
  {Astier}, {Balam}, {Balland}, {Carlberg}, {Conley}, {Fouchez}, {Guy},
  {Hardin}, {Hook}, {Howell}, {Pain}, {Perrett}, {Pritchet}, {Regnault}, \&
  {Yi}}]{Schawinski2008}
{Schawinski}, K., {Justham}, S., {Wolf}, C., {et~al.} 2008, Science, 321, 223

\bibitem[{{Schlafly} \& {Finkbeiner}(2011)}]{Schlafly2011}
{Schlafly}, E.~F. \& {Finkbeiner}, D.~P. 2011, \apj, 737, 103

\bibitem[{{Schmidt} {et~al.}(1993){Schmidt}, {Kirshner}, {Schild},
  {Leibundgut}, {Jeffery}, {Willner}, {Peletier}, {Zabludoff}, {Phillips},
  {Suntzeff}, {Hamuy}, {Wells}, {Smith}, {Baldwin}, {Weller}, {Navarette},
  {Gonzalez}, {Filippenko}, {Shields}, {Steidel}, {Perlmutter}, {Pennypacker},
  {Smith}, {Porter}, {Boroson}, {Stathakis}, {Cannon}, {Peters}, {Horine},
  {Freeman}, {Womble}, {Stone}, {Marschall}, {Phillips}, {Saha}, \&
  {Bond}}]{Schmidt1993}
{Schmidt}, B.~P., {Kirshner}, R.~P., {Schild}, R., {et~al.} 1993, \aj, 105,
  2236

\bibitem[{{Shappee} {et~al.}(2013){Shappee}, {Kochanek}, {Stanek}, {Basu},
  {Holoien}, {Jencson}, {Beacom}, {Prieto}, {Szczygiel}, {Pojmanski},
  {Dubberley}, {Elphick}, {Foale}, {Hawkins}, {Mullens}, {Rosing}, {Ross},
  {Walker}, \& {Brimacombe}}]{2013ATel5237}
{Shappee}, B.~J., {Kochanek}, C.~S., {Stanek}, K.~Z., {et~al.} 2013, The
  Astronomer's Telegram, 5237, 1

\bibitem[{{Smartt} {et~al.}(2009){Smartt}, {Eldridge}, {Crockett}, \&
  {Maund}}]{Smartt2009a}
{Smartt}, S.~J., {Eldridge}, J.~J., {Crockett}, R.~M., \& {Maund}, J.~R. 2009,
  \mnras, 395, 1409

\bibitem[{{Smartt} {et~al.}(2014){Smartt}, {Valenti}, {Fraser}, {Inserra},
  {Young}, {Sullivan}, {Pastorello}, {Benetti}, {Gal-Yam}, {Knapic},
  {Molinaro}, {Smareglia}, {Smith}, {Taubenberger}, {Yaron}, {Anderson},
  {Ashall}, {Balland}, {Baltay}, {Barbarino}, {Bauer}, {Baumont}, {Bersier},
  {Blagorodnova}, {Bongard}, {Botticella}, {Bufano}, {Bulla}, {Cappellaro},
  {Campbell}, {Cellier-Holzem}, {Chen}, {Childress}, {Clocchiatti},
  {Contreras}, {Dall Ora}, {Danziger}, {de Jaeger}, {Della Valle}, {Dennefeld},
  {Elias-Rosa}, {Elman}, {Feindt}, {Fleury}, {Gall}, {Gonzalez-Gaitan},
  {Galbany}, {Greggio}, {Guillou}, {Hachinger}, {Hadjiyska}, {Hage},
  {Hillebrandt}, {Hodgkin}, {Hsiao}, {James}, {Jerkstrand}, {Kangas},
  {Kankare}, {Kotak}, {Kromer}, {Kuncarayakti}, {Leloudas}, {Lundqvist},
  {Hook}, {Maguire}, {Manulis}, {Margheim}, {Mattila}, {Maund}, {Mazzali},
  {McCrum}, {McKinnon}, {Moreno-Raya}, {Nicholl}, {Nugent}, {Pain}, {Phillips},
  {Pignata}, {Polshaw}, {Pumo}, {Rabinowitz}, {Reilly}, {Romero-Canizales},
  {Scalzo}, {Schmidt}, {Schulze}, {Sim}, {Sollerman}, {Taddia}, {Tartaglia},
  {Terreran}, {Tomasella}, {Turatto}, {Walker}, {Walton}, {Wyrzykowski},
  {Yuan}, \& {Zampieri}}]{Smartt2014}
{Smartt}, S.~J., {Valenti}, S., {Fraser}, M., {et~al.} 2014, ArXiv:1411.0299

\bibitem[{{Soderberg} {et~al.}(2008){Soderberg}, {Berger}, {Page}, {Schady},
  {Parrent}, {Pooley}, {Wang}, {Ofek}, {Cucchiara}, {Rau}, {Waxman}, {Simon},
  {Bock}, {Milne}, {Page}, {Barentine}, {Barthelmy}, {Beardmore}, {Bietenholz},
  {Brown}, {Burrows}, {Burrows}, {Byrngelson}, {Cenko}, {Chandra}, {Cummings},
  {Fox}, {Gal-Yam}, {Gehrels}, {Immler}, {Kasliwal}, {Kong}, {Krimm},
  {Kulkarni}, {Maccarone}, {M{\'e}sz{\'a}ros}, {Nakar}, {O'Brien}, {Overzier},
  {de Pasquale}, {Racusin}, {Rea}, \& {York}}]{Soderberg2008}
{Soderberg}, A.~M., {Berger}, E., {Page}, K.~L., {et~al.} 2008, \nat, 453, 469

\bibitem[{{Spiro} {et~al.}(2014){Spiro}, {Pastorello}, {Pumo}, {Zampieri},
  {Turatto}, {Smartt}, {Benetti}, {Cappellaro}, {Valenti}, {Agnoletto},
  {Altavilla}, {Aoki}, {Brocato}, {Corsini}, {Di Cianno}, {Elias-Rosa},
  {Hamuy}, {Enya}, {Fiaschi}, {Folatelli}, {Desidera}, {Harutyunyan}, {Howell},
  {Kawka}, {Kobayashi}, {Leibundgut}, {Minezaki}, {Navasardyan}, {Nomoto},
  {Mattila}, {Pietrinferni}, {Pignata}, {Raimondo}, {Salvo}, {Schmidt},
  {Sollerman}, {Spyromilio}, {Taubenberger}, {Valentini}, {Vennes}, \&
  {Yoshii}}]{Spiro2014}
{Spiro}, S., {Pastorello}, A., {Pumo}, M.~L., {et~al.} 2014, \mnras

\bibitem[{{Swartz} {et~al.}(1991){Swartz}, {Wheeler}, \&
  {Harkness}}]{Swartz1991}
{Swartz}, D.~A., {Wheeler}, J.~C., \& {Harkness}, R.~P. 1991, \apj, 374, 266

\bibitem[{{Taddia} {et~al.}(2013){Taddia}, {Stritzinger}, {Sollerman},
  {Phillips}, {Anderson}, {Boldt}, {Campillay}, {Castell{\'o}n}, {Contreras},
  {Folatelli}, {Hamuy}, {Heinrich-Josties}, {Krzeminski}, {Morrell}, {Burns},
  {Freedman}, {Madore}, {Persson}, \& {Suntzeff}}]{Taddia2013}
{Taddia}, F., {Stritzinger}, M.~D., {Sollerman}, J., {et~al.} 2013, \aap, 555,
  A10

\bibitem[{{Tak{\'a}ts} \& {Vink{\'o}}(2006)}]{Takats2006}
{Tak{\'a}ts}, K. \& {Vink{\'o}}, J. 2006, \mnras, 372, 1735

\bibitem[{{Thompson}(1982)}]{Thompson1982}
{Thompson}, L.~A. 1982, \apjl, 257, L63

\bibitem[{{Tomasella} {et~al.}(2013){Tomasella}, {Cappellaro}, {Fraser},
  {Pumo}, {Pastorello}, {Pignata}, {Benetti}, {Bufano}, {Dennefeld},
  {Harutyunyan}, {Iijima}, {Jerkstrand}, {Kankare}, {Kotak}, {Magill},
  {Nascimbeni}, {Ochner}, {Siviero}, {Smartt}, {Sollerman}, {Stanishev},
  {Taddia}, {Taubenberger}, {Turatto}, {Valenti}, {Wright}, \&
  {Zampieri}}]{Tomasella2013}
{Tomasella}, L., {Cappellaro}, E., {Fraser}, M., {et~al.} 2013, \mnras, 434,
  1636

\bibitem[{{Tominaga} {et~al.}(2009){Tominaga}, {Blinnikov}, {Baklanov},
  {Morokuma}, {Nomoto}, \& {Suzuki}}]{Tominaga2009}
{Tominaga}, N., {Blinnikov}, S., {Baklanov}, P., {et~al.} 2009, \apjl, 705, L10

\bibitem[{{Trondal} {et~al.}(1999){Trondal}, {Granslo}, {Kushida}, {Nakano},
  {Yoshida}, \& {Kadota}}]{1999IAUC7334}
{Trondal}, O., {Granslo}, B.~H., {Kushida}, R., {et~al.} 1999, \iaucirc, 7334,
  3

\bibitem[{{Tsvetkov} {et~al.}(2009){Tsvetkov}, {Volkov}, {Baklanov},
  {Blinnikov}, \& {Tuchin}}]{Tsvetkov2009}
{Tsvetkov}, D.~Y., {Volkov}, I.~M., {Baklanov}, P., {Blinnikov}, S., \&
  {Tuchin}, O. 2009, Peremennye Zvezdy, 29, 2

\bibitem[{{Tsvetkov} {et~al.}(2006){Tsvetkov}, {Volnova}, {Shulga}, {Korotkiy},
  {Elmhamdi}, {Danziger}, \& {Ereshko}}]{Tsvetkov2006}
{Tsvetkov}, D.~Y., {Volnova}, A.~A., {Shulga}, A.~P., {et~al.} 2006, \aap, 460,
  769

\bibitem[{Turatto {et~al.}(2003)Turatto, Benetti, \& Cappellaro}]{Turatto2003c}
Turatto, M., Benetti, S., \& Cappellaro, E. 2003

\bibitem[{{Tweedy} {et~al.}(1993){Tweedy}, {Balonek}, {Wells}, {Appleton},
  {Eitter}, {Fick}, {Mikolajewski}, {Ruminski}, {Wikierski}, \&
  {Prugniel}}]{1993IAUC5769}
{Tweedy}, R., {Balonek}, T., {Wells}, L., {et~al.} 1993, \iaucirc, 5769, 1

\bibitem[{{Uomoto} \& {Kirshner}(1986)}]{Uomoto1986}
{Uomoto}, A. \& {Kirshner}, R.~P. 1986, \apj, 308, 685

\bibitem[{{Utrobin}(2007)}]{Utrobin2007}
{Utrobin}, V.~P. 2007, \aap, 461, 233

\bibitem[{{Utrobin} \& {Chugai}(2008)}]{Utrobin2008}
{Utrobin}, V.~P. \& {Chugai}, N.~N. 2008, \aap, 491, 507

\bibitem[{{Utrobin} \& {Chugai}(2009)}]{Utrobin2009}
{Utrobin}, V.~P. \& {Chugai}, N.~N. 2009, \aap, 506, 829

\bibitem[{{Valenti} {et~al.}(2014){Valenti}, {Sand}, {Pastorello}, {Graham},
  {Howell}, {Parrent}, {Tomasella}, {Ochner}, {Fraser}, {Benetti}, {Yuan},
  {Smartt}, {Maund}, {Arcavi}, {Gal-Yam}, {Inserra}, \& {Young}}]{Valenti2014b}
{Valenti}, S., {Sand}, D., {Pastorello}, A., {et~al.} 2014, \mnras, 438, L101

\bibitem[{{van Dokkum}(2001)}]{vanDokkum2001}
{van Dokkum}, P.~G. 2001, \pasp, 113, 1420

\bibitem[{{Yamaoka} {et~al.}(2004){Yamaoka}, {Itagaki}, {Klotz}, {Pollas}, \&
  {Boer}}]{2004IAUC8413}
{Yamaoka}, H., {Itagaki}, K., {Klotz}, A., {Pollas}, C., \& {Boer}, M. 2004,
  \iaucirc, 8413, 2

\bibitem[{{Young}(2004)}]{Young2004}
{Young}, T.~R. 2004, \apj, 617, 1233

\bibitem[{{Zampieri} {et~al.}(2003){Zampieri}, {Pastorello}, {Turatto},
  {Cappellaro}, {Benetti}, {Altavilla}, {Mazzali}, \& {Hamuy}}]{Zampieri2003}
{Zampieri}, L., {Pastorello}, A., {Turatto}, M., {et~al.} 2003, \mnras, 338,
  711

\end{thebibliography}

\appendix

\section{Light curve models}
\label{section:Light_curve_models}

In the following we summarize a few basic concepts that help us understand the observed trends in our sample of Type II-P/L SNe as discussed in Section \ref{section:Results_and_discussion}.
In this chapter we first have a brief look into the different phases of a light curve (Section \ref{section:Light_curve_phases}) and give an overview over two analytical models for the early light curve phase by \citet{Arnett1980,Arnett1982} and \citet{Rabinak2011} (Section \ref{section:Analytical_models_for_cooling_phase}). In Section \ref{section:Parameter_dependencies_in_the_models}, finally, we investigate the various parameter dependencies in the models.

\subsection{Light curve phases}
\label{section:Light_curve_phases}

Radiation hydrodynamic calculations of the explosion and evolution of Type II SNe
have been presented by \citet{Grassberg1971}, \citet{Falk1977}, \citet{Falk1978}, \citet{Klein1978}, \citet{Hillebrandt1981}, and \citet{Litvinova1983,Litvinova1985}, as well as more specifically for SNe II-P by \citet{Young2004}, \citet{Utrobin2007}, \citet{Utrobin2009}, \citet{Bersten2011}, and \citet{Dessart2011}, for SNe II-L by \citet{Swartz1991} and \citet{Blinnikov1993} and for the peculiar Type II-P SN 1987A by \citet{Blinnikov2000b}, \citet{Dessart2010} and \citet{Pumo2011}. Here we give a summary view of the physical processes at play and the various evolutionary stages seen in these simulations.

The light curve of a Type II SN can be divided into four distinct phases depending on the physical processes at play; the shock breakout phase, the cooling phase, the recombination phase, and the nebular phase \citep[e.g.][]{Blinnikov1993,Young2004}. The first phase is the shock breakout phase, where the internal energy of the shock diffuses out to give an X-ray (extended progenitor) or gamma-ray (compact progenitor) burst. This phase lasts only for minutes or hours \citep{Falk1977} and only been observed for two Type II-P SNe \citep[SNLS-04D2dc and SNLS-06D1jd][]{Gezari2008,Schawinski2008}) and once for a Type I SN (SN 2008D, \citealt{Soderberg2008}).

Over the next few weeks, the ejecta expand and cool. As the post-shock temperatures are $\gtrsim 10^6$\,K, hydrogen and helium are fully ionized and remain so until the temperature has decreased down to $\sim 10^4$\,K, when recombination begins. This takes a few weeks according to models \citep{Utrobin2007,Bersten2011}. Radiative diffusion is  inefficient when compared to adiabatic cooling \citep{Bersten2011}.

The ejecta are dynamically evolving during the first few days. The adiabatic cooling accelerates the layers from their initial post-shock velocity to their final coasting velocity, which is a factor of $\sqrt2$ higher (assuming equipartition between internal energy and expansion energy in the post-shock flow). The inner regions also experience strong Rayleigh-Taylor mixing that breaks spherical symmetry \citep{Herant1991,Mueller1991}.

The time scale to reach the final dynamic state is approximately the time scale of adiabatic cooling $t_{\rm ad} \sim R/v$ (for an adiabatically cooling, radiation dominated gas $E_{\mathrm{int}} \propto R^{-1}$, where $E_{\mathrm{int}}$ is the internal energy) which is 1\,day for a radius $R = 1000$\,R$_{\astrosun}$ and a velocity $v=10^4$\,km\,s$^{-1}$. Thus, unless the explosion energy is very small (giving $v \ll 10^4$\,km\,s$^{-1}$), it is a good approximation to take the ejecta as coasting for $t \gtrsim 1$\,day.

Homologous expansion ($v \propto R$) is achieved on a similar time scale, as it is fulfilled when $R \gg$ initial radius, $R_0$, the same condition for adiabatic acceleration to be complete.

When the temperature goes below $\sim 10^4$\,K, recombination of helium and hydrogen has to be taken into account. By $\sim$ 3 weeks this begins in the outermost layers of the ejecta and gradually moves into deeper layers as a recombination and cooling wave. As the internal energy can easily escape when the gas ahead of it recombines and the opacity decreases dramatically, energy is now lost by radiation. The recombination wave effectively ploughs through the envelope, releasing internal energy at a roughly constant rate giving a plateau in the light curve.

At some point in the recombination phase, gamma-rays from $^{56}$Ni located deeper inside the ejecta start to reach the envelope layers and affect the thermal state. In particluar for the later parts of the recombination phase, this must be taken into account \citep{Eastman1994,Kasen2009b}.

In order to achieve a linear decline, the recombination wave needs to be avoided or weakened. A sufficiently low density, for example due to a very low-mass or a very extended pre-supernova star might produce the desired effect \citep{Blinnikov1993}.

After $\sim$100\,days the entire ejecta have recombined and become optically thin in the continuum. Internal energy is maintained by $^{56}$Co, which heats and ionizes the gas to keep emission lines active.

\subsection{Analytical models for the cooling phase}
\label{section:Analytical_models_for_cooling_phase}

Analytical models have the advantage of easy implementation, allowance for
physical understanding, and rapid calculations over parameter-space.
Analytical models for the cooling phase ($\sim 0.5-20$\,days) have been presented by several authors \citep[e.g.][]{Arnett1980,Arnett1982,Chugai1991,Popov1993,Zampieri2003,Kasen2009b,Nakar2010,Rabinak2011,Chatzopoulos2012,Nagy2014}.
In this article we focus on two particular models; those of \citet{Arnett1980,Arnett1982} (A80, A82) and \citet{Rabinak2011} (R11).

\subsubsection{Arnett 1980, 1982}
\citet{Arnett1980,Arnett1982} presented the first analytical models for the light curves of supernovae. A80 presented models for explosion-powered light curves, whereas A82 extended these to include $^{56}$Ni (there are however also some clarifications in this paper over the use of the A80 models). This framework assumes spherical symmetry, homologous expansion, local thermodynamic equilibrium, a radiation dominated equation of state, constant opacity, and certain constraints on the density profile. The first law of thermodynamics is solved with an Eddington outer boundary condition. To obtain an analytic solution, it is postulated that the temperature profile is self-similar.

The luminosity solution for constant density (labelled ``$A=0$'' in A80\footnote{A80 also gives solutions for other density profiles, but the differences are relatively small and will not be explored further here.}) is
\begin{equation}
\label{equation:luminosity_Arnett}
L(t) = L_0(M,R_0,E,\kappa)\exp{\left[-\frac{t}{\tau_0(M,R_0,\kappa)}\left(1 + \frac{t}{{2 \tau_h(M,R_0,E)}}\right)\right]} ,
\end{equation}
with (see Equations 14, 37 and 40 in \citealt{Arnett1980} and Equations 19 and 22 in \citealt{Arnett1982})
\begin{eqnarray}
L_0(M,R_0,E,\kappa) &=& 5.2\e{43} \left(\frac{R_0}{10^{14}\,\mathrm{cm}}\right) \left(\frac{E}{2\e{51}\,\mathrm{erg}}\right) \nonumber \\
& &\times \left(\frac{\kappa}{0.4\,\mathrm{cm}^2\,\mathrm{g}^{-1}}\right)^{-1} \left(\frac{M}{M_{\astrosun}}\right)^{-1} \mbox{erg s}^{-1} ,
\end{eqnarray}
where
\begin{equation}
\tau_h(M,R_0,E) = \left(\frac{10}{3}\right)^{-1/2} R_0 E^{-1/2} M^{1/2} \mbox{s} ,
\end{equation}
and
\begin{equation}
\tau_0(M,R_0,\kappa) = \frac{1}{\beta c} \kappa M R_0^{-1} \mbox{s} ,
\end{equation}
where $\tau_0$ is the diffusion time scale at $t=0$ (and therefore has no dependency on $E$), $\tau_h$ is the initial doubling time of the radius, $\beta$ is a parameter that depends on the density profile (A80 shows that $\beta \approx 13.7$ can be used for a variety of density profiles), and $\kappa$ is the opacity. We have here used $E=\frac{3}{10} M v_{\mathrm{sc}}^2$ (the Arnett formulation is $\tau_h = R_0 / v_{\mathrm{sc}}$, where $v_{\mathrm{sc}}$ is the scaling velocity). 

For $t/\tau_h \ll 1$, the luminosity is $L(t) \approx e^{-t/\tau_0}$. As $\tau_0$ is of the order of $10^4$\,days, the bolometric light curve is initially almost flat. For $t/\tau_h \gg 1$ the luminosity declines as $e^{-t^2/(2\tau_0\tau_h)}$, where $\tau_h$ is the time required to reach a size of twice the progenitor size, which is of the order of 1 day.
The effective temperature is computed by applying the blackbody formula to $L(t)$ and $R(t) = R_0 + v_{\mathrm{sc}} t$ (see Equations 43 and 44 in A80).

\subsubsection{Rabinak \& Waxman 2011}
The R11 framework starts with a specification of the progenitor density profile, and then computes post-shock density and temperature assuming a strong, radiation-dominated shock, and a analytical shock propagation formula from \citet{Matzner1999}. The ejecta are then assumed to evolve adiabatically.

The luminosity is not found from the diffusion equation as in A80, but from determining the photospheric location assuming a constant $\kappa$, taking the temperature in this layer to equal the effective temperature, and then applying the blackbody formula. 

The resulting luminosity and temperature for a red supergiant progenitor (density power law index $n=3/2$ for efficiently convective envelopes) are (see Equations 13 and 14 in R11):
\begin{equation}
L(t) = 8.5\e{42} \frac{E_{51}^{0.92} R_{0,13}}{f_\rho^{0.27} (M/M_{\astrosun})^{0.84} \kappa^{0.92}_{0.34}}  t_5^{-0.16} \mathrm{erg}\,\mathrm{s}^{-1}
\end{equation}
and
\begin{equation}
T_{\mathrm{ph}}(t) = 1.6 f_\rho^{-0.037} \frac{E_{51}^{0.027} R_{0,13}^{1/4}}{(M/M_{\astrosun})^{0.054} \kappa^{0.28}_{0.34}} t_5^{-0.45} \mathrm{eV} ,
\end{equation}
where $E_{51}$ is the explosion energy in units of $10^{51}$\,erg, $R_{0,13}$ is the progenitor radius is units of $10^{13}$\,cm, $f_\rho$ is a parameter related to the
progenitor density profiles (we use $f_\rho$ = 0.1; \citealt{Calzavara2004} find $f_\rho$ = 0.079-0.13 for different RSG structures), $\kappa_{0.34}$ is the opacity in units of 0.34\,cm$^2$\,g$^{-1}$ and $t_5$ is the time in units of $10^5$\,s.

\subsubsection{Physical approximations}

Even though the aforementioned models help us in our understanding of the light curve morphology, there are limitations that need to be considered.
One is the assumption of constant opacity. Full ionization and pure electron scattering give $\kappa$ = 0.4\,cm$^2$\,g$^{-1}$ (pure hydrogen) and $\kappa$ = 0.34\,cm$^2$\,g$^{-1}$ (solar hydrogen and helium mixture). These are reasonable estimates at early times, although free-free absorption may contribute as well \citep{Rogers1992}. At later times recombination reduces the electron scattering opacity but line blocking
makes significant contributions \citep{Karp1977}.

As mentioned above the ejecta are dynamicaly evolving in the cooling phase, which is not considered in these models. The outer layers of relevance do, however, only accelerate by a factor of $\sqrt 2 = 1.4$ so this effect should not be strong.

Weak or moderate mixing of $^{56}$Ni has no influence for Type II-P light curves during the first weeks \citep{Kasen2009b}. However, if there would be very strong mixing, $^{56}$Ni bullets shot into the outer envelope could have an impact on the cooling phase. $^{56}$Ni becomes important earlier for more compact progenitors (e.g. for SN 1987A, with $R_0 = 50$\,R$_{\astrosun}$, the $^{56}$Ni-driven second peak began already after a week; \citealt{Blinnikov2000b}). This limits the applicability of the models to radii $\gtrsim 100$\,R$_{\astrosun}$.

There is mounting evidence that the SN explosion mechanism is asymmetric, and that the metal cores obtain highly asymmetric shapes \citep[e.g.][]{Kjaer2010,Larsson2013,Grefenstette2014}. However, the shocks tends to spherize at larger mass coordinate, and the outer layers of interest here are likely close to spherical symmetry \citep[e.g.][]{Leonard2001,Leonard2006}.

A radiation dominated equation of state should be an excellent approximation for these early phases (A82). 
The R11 models have more realistic density profiles
and photospheric position than A80, on the other hand they do not include
radiative diffusion which A80 does.

\subsection{Parameter dependencies in the models}
\label{section:Parameter_dependencies_in_the_models}

Here we discuss the influence of model parameters $M$, $E$, and $R_0$ on light curve properties (bolometric and optical rise times and peak luminosities).

\subsubsection{Radius}
\label{section:Radius}

The dependency of the bolometric luminosity on the progenitor radius is $L \propto R_0$ both in the A80 and the R11 model. A larger radius leads to higher luminosity because adiabatic losses are smaller and more of the explosion energy is retained at any given time.

Both in A80 and R11, larger $R_0$ also leads to brighter optical peaks, driven by the dependency of $L_{\mathrm{bol}}$. A larger $R_0$ additionally means it takes longer to rise to optical peak, which counteracts this, but not enough to reverse the trend.

The optical light curve can be thought of as the bolometric light curve multiplied by the fraction of the bolometric light that emerges in the optical:
\begin{equation}
L_{\mathrm{opt}}(t) = L_{\mathrm{bol}}(t) \times f_{\mathrm{opt}}(t) .
\end{equation}
The first function we have explicit equations for, the second is computed from the photospheric temperature. The optical peak will be the peak of $f_{\mathrm{opt}}(t)$, shifted to a somewhat earlier time by the monotonically decreasing $L_{\mathrm{bol}}(t)$ function. Under the assumption of constant opacity, the $f_{\mathrm{opt}}$ function depends only on the photospheric temperature, which in turn depends on $L_{\mathrm{bol}}(t)$ and $R_{\mathrm{phot}}(t)$. For $R \gg R_0$, $f_{\mathrm{opt}}(t)$ depends on $R_0$ only through its influence on $L_{\mathrm{bol}}$. 

As exemplified by the R11 models, the bolometric luminosity follows $L(T) \sim t_5^{-0.16}$ and changes on a time scale of $\tau_L = L/\mathrm{d}L/\mathrm{d}t = 6.3\,t$. For our typical rise times of 5-15 days, the bolometric luminosity thus changes on a time scale of 32-95 days. The temperature instead changes on a faster time scale $\tau_T = T_{ph}/\mathrm{d}T_{ph}/\mathrm{d}t = 2.2\,t$ corresponding to 11-33 days for our typical rise times. The time scales in the A80 formalism show similar relations. This means that the optical light curve peak must lie close in time to the peak in $f_{\mathrm{opt}}$.

\begin{figure}
  \centering
    \includegraphics[width=9cm]{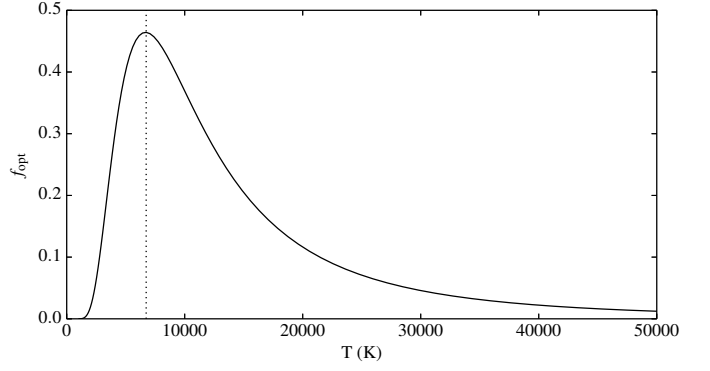}
    \caption{Fraction of the of the bolometric light, $L_{\mathrm{opt}}$, that emerges in the optical (4000-8000\,\AA). $f_{\mathrm{opt}}$ peaks when $T_{\mathrm{phot}} \approx 7000$\,K (dotted line).}
    \label{figure:Anders}
\end{figure}

We define optical as the 4000-8000\,\AA\ range. Figure \ref{figure:Anders} shows 
$f_{\mathrm{opt}}$ for this range as a function of $T$. $f_{\mathrm{opt}}$ peaks when $T_{\mathrm{phot}} \approx 7000$\,K. Indeed, we find that for the A80/A82 model the rise to optical peak mostly ends at temperatures between 6400 and 7200\,K, and for the R11 model at temperatures between 7500 and 7900\,K for a range of parameters (see Section \ref{section:Rise_times} for the exact parameter space).

We further assume the peak of $L_{\mathrm{opt}}$ is close to the peak of $f_{\mathrm{opt}}$, and $R \gg R_0$. Then
\begin{equation}
T_{\mathrm{phot}} = \left(\frac{L_{\mathrm{bol}}(t)}{4\pi \sigma \left(v_{\mathrm{phot}}t\right)^2}\right)^{1/4} ,
\end{equation}
where $v_{\mathrm{phot}}$ is the photospheric velocity and $\sigma$ is the Stefan-Boltzmann constant. As $L_{\mathrm{bol}} = C R_0 E M^{-1}\exp{\left(-t^2/(2\tau_0\tau_{\mathrm{h}})\right)}$ for $t \gg \tau_{\mathrm{h}}$
\citep{Arnett1980} and $v_{\mathrm{phot}}^2 \approx 10/3 E M^{-1}$, we get for $t \ll 2\tau_0\tau_{\mathrm{h}}$
\begin{equation}
T_{\mathrm{phot}}(t) \approx \frac{C^{1/4} R_0^{1/4}}{\left(\frac{40\pi}{3} \sigma\right)^{1/4}t^{1/2}} ,
\end{equation}
where $C=5.2\e{11}$\,g\,cm$^{-1}$\,s$^{-1}$ for $\kappa=0.4$ cm$^2$ g$^{-1}$. It thus takes longer to reach 7000\,K (or any other temperature) for a larger $R_0$ because $R_0$ increases the scale of $L_{\mathrm{bol}}$. The dependency on $M$ and $E$ drops out, as $L_{\mathrm{bol}}$ and $v_{\mathrm{phot}}^2$ both depend in the same way on them ($\propto E/M$); in this way we can understand why $E$ and $M$ are not strong drivers of the optical rise times in the models (see Figures \ref{figure:model_data_risetime_absmag_comparison_Arnett} and \ref{figure:model_data_risetime_absmag_comparison_Rabinak} in Section \ref{section:Results_and_discussion}).

We can solve this Equation for the rise time:
\begin{equation}
t_{\mathrm{opt-rise}} \approx \frac{C^{1/2} R_0^{1/2}}{ \left(\frac{40 \pi}{3}\sigma\right)^{1/2}T_{\mathrm{peak}}^2} ,
\end{equation}
and we get
\begin{equation}
\label{equation:t_optrise_Arnett}
t_{\mathrm{opt-rise}} \approx 9.2\,\mathrm{d} \left(\frac{R_0}{100~R_{\astrosun}}\right)^{1/2}\left(\frac{T_{\mathrm{peak}}}{7000 K}\right)^{-2} . 
\end{equation}
The R11 model gives (solving $T_{\mathrm{phot}}(t) = T_{\mathrm{peak}}$):
\begin{equation}
\label{equation:t_optrise_Rabinak}
t_{\mathrm{opt-rise}} \approx 6.8\,\mathrm{d} \left(\frac{R_0}{100~R_{\astrosun}}\right)^{0.56} \left(\frac{T_{\mathrm{peak}}}{7000 K}\right)^{-2.2} E_{51}^{0.06} \left(\frac{M}{10 M_{\astrosun}}\right)^{-0.12} . 
\end{equation}
The dependencies are similar to the ones derived for A80; we note the weak dependencies on $E$ and $M$.

With these formulas we can understand why redder bands peak later; they have lower $T_{\mathrm{peak}}$. We now also have a quantitative handle on the dependency on $R_0$; it is a square root dependency which fits reasonably well with the full solutions (see Figures \ref{figure:model_data_risetime_absmag_comparison_Arnett} and \ref{figure:model_data_risetime_absmag_comparison_Rabinak} in Section \ref{section:Results_and_discussion}). 
Using hydrodynamical simulations \citet{Swartz1991} also find that larger progenitor radii result in supernovae with longer rise times and brighter optical peaks ($B$-band). 
However, as these authors discuss, significant He enrichment of the envelope could have a similar effect to smaller envelope masses in that the light curve will peak earlier, be brighter, and still have a linear decline. The reason is that helium has a lower opacity than hydrogen and therefore the energy trapping is less effective. This means that in principle the observed brightness for SNe II-L might also be explained by a He-rich envelope. In our parameter study we assume a constant opacity, $\kappa$, and therefore cannot account for such effects. 

\subsubsection{Mass and explosion energy}
The dependency of the bolometric luminosity on mass is $L \propto M^{-1}$ and $L \propto M^{-0.84}$ in the A80 and R11 models, respectively. A higher mass reduces the peak luminosity as it is more difficult for photons to diffuse out of more massive ejecta. 
\citet{Swartz1991} also find that higher mass envelopes result in supernovae with dimmer optical peaks ($B$-band). 

The dependency of the bolometric luminosity on energy is $L \propto E$ in the A80 model, and $L \propto E^{0.92}$ in the R11 model. A higher energy leads to higher peak luminosity for two reasons; more internal energy is created in the ejecta, and it can diffuse out more rapidly as higher $E$ leads to higher velocities and lower density. On the other hand, higher velocities also lead to stronger adiabatic losses, so the final outcome is not obvious. The analytic solutions show that the first effects dominate.

If we ignore the first day or so when the progenitor size affects the time scales, the diffusion time in A80 is
\begin{equation}
\sqrt{\tau_0\tau_{\mathrm{h}}} \propto M^{3/4} E^{-1/4} .
\end{equation}
Higher mass leads to longer light curves, as it takes longer for the energy to diffuse out. Energy has a relatively weak influence, but higher energy leads to faster light curves as photon escape and adiabatic degrading occur faster.

Higher mass also leads to dimmer optical peaks (both models). Higher $M$ scales down $L_{\mathrm{bol}}$ (see above), and this typically has a stronger impact than the slower decline rate obtained from a longer $\tau_0\tau_{\mathrm{h}}$.
A higher explosion energy, instead, gives brighter optical peaks. This is driven by the influence of $E$ on $L_{\mathrm{bol}}$, as the evolution of $T_{\mathrm{phot}}$ only weakly depends on $E$.

Regarding rise time, A80 gives somewhat longer rise times with higher $M$ and lower $E$ (the weak dependency is absent in our approximate formula \ref{equation:t_optrise_Arnett}), whereas R11 gives shorter ones. One difference between these models is that the photosphere is fixed in mass coordinate in A80, but moves inwards in R11. 
In A80 a lower $E/M$ gives higher $\tau_0\tau_{\mathrm{h}}$ which means it takes longer to reach $T_{\mathrm{peak}}$ (through the exponential factor in Equation \ref{equation:luminosity_Arnett}). The dependence of the rise time on mass and explosion energy, however, is smaller than the dependence on the radius, as we can see in Equation \ref{equation:t_optrise_Arnett}. In fact our estimate for the rise time shows that to first order, the effects of mass and energy on the rise time is negligible.

\section{Additional tables}

Table \ref{table:sequence_stars} lists the local sequence stars from the SDSS DR9 catalogue that were used to calibrate the LT and LCOGT $g'r'i'$-band photometry of LSQ13cuw to the SDSS system (see also Section \ref{section:Data_reduction}).

\begin{table*}
 \caption{Optical sequence stars}
 \label{table:sequence_stars}
 \centering
 \begin{tabular}{c c c		       c        c       c        c       c        c       }
 \hline                 
 Star & R.A.        & Dec.        & mag    & error & mag    & error & mag    & error \\ 
      &             &             & $g'$   &       & $r'$   &       & $i'$   &       \\
\hline                 
1     & 02:39:41.21	& -08:28:16.6 &	18.640 & 0.008 & 17.191	& 0.005	& 16.404 & 0.005 \\
2     & 02:39:43.66	& -08:30:21.2 &	18.794 & 0.009 & 18.505	& 0.009	& 18.416 & 0.011 \\
3     & 02:39:50.81	& -08:28:14.9 &	18.112 & 0.006 & 16.681	& 0.005	& 15.869 & 0.004 \\
4     & 02:39:45.81	& -08:33:40.3 &	17.402 & 0.005 & 17.020	& 0.005	& 16.879 & 0.005 \\
5     & 02:39:55.97	& -08:28:56.6 &	17.287 & 0.005 & 16.312	& 0.004	& 15.961 & 0.004 \\
6     & 02:40:02.10	& -08:29:36.2 &	17.280 & 0.005 & 16.814	& 0.005	& 16.632 & 0.005 \\
7     & 02:40:02.22	& -08:30:06.9 &	18.213 & 0.007 & 17.193	& 0.005	& 16.825 & 0.005 \\
8     & 02:40:02.88	& -08:30:34.8 &	18.180 & 0.007 & 17.160	& 0.005	& 16.779 & 0.005 \\
9     & 02:40:05.63	& -08:28:06.6 &	17.945 & 0.006 & 16.494	& 0.004	& 15.695 & 0.004 \\
10    & 02:39:57.24	& -08:34:40.3 &	19.135 & 0.011 & 18.255	& 0.008	& 17.893 & 0.008 \\
11    & 02:40:08.63	& -08:33:27.7 &	16.301 & 0.004 & 15.425	& 0.004	& 15.140 & 0.004 \\
12    & 02:40:12.25	& -08:32:48.9 &	16.952 & 0.004 & 16.480	& 0.004	& 16.285 & 0.004 \\
13    & 02:40:05.99	& -08:28:47.7 &	19.115 & 0.010 & 17.686	& 0.006	& 17.053 & 0.006 \\
 \hline  
 \end{tabular}
 \\[1.5ex]
 \flushleft
  Coordinates and magnitudes as reported by SDSS.
\end{table*}

Table \ref{table:IIbs} shows $r'/R$-band properties of SNe that were not included in our II-P/L sample, but are shown in Figures \ref{figure:absmag_vs_rise_nolabels} and \ref{figure:model_data_risetime_absmag_comparison} for comparison.

\begin{table*}
  \caption{$r'/R$-band properties of comparison SNe}
  \label{table:IIbs}
  \centering
  \begin{tabular}{l c          l                              l                          c                      r@{}         l                          c                      }  
  \hline                      %
  SN              & Type      & \multicolumn{2}{c}{$E(B - V)$}                            & End of rise          & \multicolumn{2}{c}{Rise time in}    &                       \\ 
                  &           & \multicolumn{1}{c}{Galactic} & \multicolumn{1}{c}{Host} & absolute magnitude   & \multicolumn{2}{c}{days (rest frame)} &                       \\ 
 \hline                       %
  SN 1987A        & II-P pec  & 0.066                        &  -                       & $-$14.65 $\pm$ 0.29  &  10.2      & $\ \pm$ 1.0              & a,b,c                 \\ 
  SN 2009kr       & II-L      & 0.07                         &  -                       & $-$16.86 $\pm$ 0.15  & $\geq$ 5.0 &                          & a,b,d,e               \\ 
  SN 1993J        & IIb       & 0.187                        &  -                       & $-$17.87 $\pm$ 0.44  &  22.3      & $\ \pm$ 0.9              & a,b,f,g,h,i,j,k,l,m,n \\ 
  SN 2008ax       & IIb       & 0.022                        &  -                       & $-$17.63 $\pm$ 0.30  &  21.6      & $\ \pm$ 0.3              & a,b,o,p               \\ 
  SN 2011dh       & IIb       & 0.031                        &  -                       & $-$17.39 $\pm$ 0.01  &  20.0      & $\ \pm$ 0.5              & a,b,q                 \\ 
  \hline  &      
  \end{tabular}
  \\[1.5ex]
  \flushleft
a) this paper;
b) NASA/IPAC Extragalactic Database;   
c) \citealt{Hamuy1988};      
d) \citealt{2009CBET2006};   
e) \citealt{Elias-Rosa2010}; 
f) \citealt{1993IAUC5731};   
g) \citealt{1993IAUC5742};   
h) \citealt{1993IAUC5761};   
i) \citealt{1993IAUC5765};   
j) \citealt{1993IAUC5769};   
k) \citealt{1993IAUC5774};   
l) \citealt{1993IAUC5832};   
m) \citealt{Lewis1994};      
n) \citealt{Barbon1995};     
o) \citealt{Pastorello2008}; 
p) \citealt{Tsvetkov2009};   
q) \citealt{Ergon2014a};	 
\end{table*}

\end{document}